\documentclass[
reprint, floatfix,
 amsmath,amssymb,
 aps,
 pre
]{revtex4-2}
\usepackage{dcolumn}
\usepackage{bm}
\usepackage{amsmath}
\usepackage{mathtools}
\usepackage{xcolor}
\usepackage{cancel}
\usepackage{comment}

\usepackage{xcolor} 
\usepackage[normalem]{ulem} 
\newcommand\hl{\bgroup\markoverwith
  {\textcolor{yellow}{\rule[-.5ex]{2pt}{2.5ex}}}\ULon}

\usepackage[normalem]{ulem}

\newcommand{\FokkerPlanck}{{\mathcal{L}}_{\lambda}}

\newcommand{\eq}{\mathrm{eq}}
\newcommand{\peq}{\rhoeq_{\lambda}}
\newcommand{\potentialCommand}{V_{\lambda}}
\newcommand{\neqcorrection}{\delta \rho_t}

\newcommand{\rhoeq}{\rho^\mathrm{eq}}

\newcommand{\boldx}{x}

\newcommand{\dt}{{\mathrm{d}t}}
\renewcommand{\d}{{\mathrm{d}}}
\newcommand{\dx}{{\mathrm{d}\boldx}}

\begin{document}

\preprint{APS/123-QED}

\title{
Higher-order response theory in optimal stochastic thermodynamics
}
\author{Sam D'Ambrosia$^{1,2}$}
\email{shda@berkeley.edu}
\author{Adrianne Zhong$^{1,2,3}$}
\author{Michael R. DeWeese$^{1,2,4}$}
\affiliation{%
${^1}$Department of Physics, University of California, Berkeley, Berkeley, CA, 94720 \\
${^2}$Redwood Center For Theoretical Neuroscience, University of California, Berkeley, Berkeley, CA, 94720 \\
${^3}$NSF-Simons National Institute for Theory and Mathematics in Biology, Chicago, IL, 60611\\
${^4}$Department of Neuroscience, University of California, Berkeley, Berkeley, CA, 94720
}%
\begin{abstract}
    Linear response theory has found many applications in statistical physics. One of these is to compute minimal-work protocols that drive nonequilibrium systems between different thermodynamic states, which are useful for designing engineered nanoscale systems and understanding biomolecular machines. We clarify the relationships between linear-response-based approximations used to study optimal protocols in different driving regimes by demonstrating how they arise from a truncation of a causal response, or Volterra, series expansion. We then construct higher-order response terms and discuss the utility and drawbacks of their inclusion in the determination of work-minimizing protocols. We focus on the cases of an overdamped particle in a harmonic trap, and a double well system modeled as a two-state discrete system, ultimately showing that while the inclusion of higher-order response in work approximations can improve predictions of excess work and modestly improve the resulting higher-order optimal protocols, a significant computational expense is incurred while introducing the possibility of predicting arbitrarily low and unphysical negative excess work. 
\end{abstract}

\date{\today}

\maketitle

\section{Introduction}

A central goal of nonequilibrium thermodynamics is to characterize and calculate optimal, minimal-work protocols that transform a thermodynamic system between different states in finite time \cite{seifert2007, blaber2023optimal}. These protocols find applications in fields such as the engineering of nanoscopic devices \cite{diana2013finite, zulkowski2014optimal, proesmans2020finite, proesmans2020optimal, whitelam2023train, schmiedl2007efficiency, martinez2017colloidal, martin2018extracting, abiuso2020optimal, brandner2020thermodynamic, frim2022optimal, frim2022geometric}, and in understanding biophysical systems \cite{geiger2010optimum, dellago2013computing, sivak2016thermodynamic, lucero2019optimal,  blaber2022efficient, davis2024active, wareham2024multi}. However, exact, analytic optimal protocols exist only for the simplest systems, such as a Brownian particle in a harmonic potential \cite{seifert2007}. Alternatively, a wide range of approximate results rely on assumptions based on linear response, which assumes that the thermodynamic system remains close to the driving protocol's equilibrium state at all times~\cite{kubo1957, zwanzig2001nonequilibrium, limmer2022statmech}. These include the range of results from slow driving thermodynamic geometry \cite{sivak2012, zulkowski2012geometry, rotskoff2015optimal, rotskoff2017geometric, zulkowski2015optimal, frim2022geometric, frim2022optimal}, ansatz-based slow driving results \cite{bonanca2014}, and results applying in the weak driving regime \cite{bonanca2018, naze2022optimal, naze2024analytical}. 

However, linear response corresponds only to the first-order term of a more general response expansion, or Volterra series. While the Volterra series is well known in control theory \cite{boyd1984analytical, BROCKETT1976167, wiener1942response, FRISTON20031273, lesiak1977}, in physics it often appears implicitly such as in time-ordered response \cite{kubo1957, zwanzig2001nonequilibrium, limmer2022statmech} or Dyson series \cite{Dyson1949}. Here, we use it as a clarifying framework to organize and compare existing linear-response approximations and to explore higher-order control in the context of stochastic thermodynamics. 

Nonlinear response has been studied in nonequilibrium stochastic settings, including through results for stochastic spin systems \cite{Lippiello_2008}, through connections to correlations with entropy production fluctuation–dissipation relations \cite{Holsten_2021}, and using path integrals \cite{Baiesi_Maes_2009, Basu2014AO, Colangeli_2011}. Expansions of the excess work based on nonlinear response theory have also been investigated for thermally isolated Hamiltonian systems, which considered the case of control by a linear protocol \cite{Naz2022SeriesEO}. Here, we focus on utilizing nonlinear response to obtain work-minimizing control protocols for open stochastic systems.

Our goal in this manuscript is both to clarify the basis of, and relationships among linear response approximations as used in optimal stochastic thermodynamics, and also to study higher order response in this context. First, we demonstrate how existing linear response approaches to optimal stochastic thermodynamics such as thermodynamic geometry and ansatz-based weak driving are derived from the first term of the Volterra series. We also demonstrate the equivalence of linear-response thermodynamic geometry with the time-local approach from \cite{Wadia_2022} and \cite{sawchuk2024dynamical}, and extend this time-local approximation to all orders, observing both that (1) the inclusion of higher orders in the excess work cost function can lead to predictions of unphysical negative excess work, and that (2) jumps found in true optimal protocols \cite{seifert2007, zhong2024, aurell2011optimal, Aurell2012BoundaryLayers} can cause the approximation to diverge. 

Following this, we construct higher-order response terms to arbitrary order in the context of finite-time thermodynamics, and evaluate the utility of using higher-order corrections beyond linear response to derive optimal finite-time protocols, demonstrating results for an overdamped harmonic oscillator and a double-well system modeled as a two-state discrete 
system. 
We find that while including higher-order response terms can lead to more accurate 
predictions for minimal excess work and marginally improve the performance of the derived optimal protocols, their inclusion comes at a substantial computational cost, and again allows for the possibility of predicting unphysical negative excess work. 

These obstacles further motivate a different but related approach to optimal stochastic thermodynamics based on a recently discovered connection between optimal transport and thermodynamic geometry \cite{zhong2024, ito2024geometric}, which avoids relying altogether on limited near-equilibrium expansions. We give a brief discussion of this in Section IV.

\subsection*{Preliminaries}

The results of our manuscript are valid for passive (i.e., detailed-balance preserving) overdamped and underdamped Langevin dynamics, as well as passive discrete state Markov dynamics. 
Given a Hamiltonian $H_\lambda(x)$ parameterized by control-parameters $\lambda \in \mathcal{M}$, the expected work needed to implement a finite-time protocol $\Lambda =\lambda(t)|_{t \in [0, t_f]}$ connecting $\lambda(0) = \lambda_i$ to $\lambda(t_f) = \lambda_f$ is given by 
\begin{equation}
    \langle W \rangle = \int_0^{t_f} \dot{\lambda}^a \bigg\langle \frac{\partial H_\lambda}{\partial \lambda^a} \bigg\rangle \, \dt,\label{eq:total_work}
\end{equation}
where we have suppressed the explicit time dependence of $\lambda$, $H$, and the average over the nonequilibrium distribution. Here we focus on the continuous case where $x \in \mathbb{R}^d$. For the discrete case $x \in \Omega$ where $\Omega$ represents a finite discrete state space, integrals over $x \in \mathbb{R}^d$ become sums over $x \in \Omega$ (e.g., as specified in \cite{sawchuk2024dynamical}).

Here, $\mathcal{M}$ is the 
$k$-dimensional control manifold. Throughout this paper we utilize the Einstein summation convention where repeated upper and lower indices denote implicit summation, and the dot denotes a total time derivative $\dot{\lambda}^a := \d \lambda^a / \dt$. Here we will use Latin indices, with control parameters in specific examples referred to with Greek letters. The unadorned brackets $\langle \cdot \rangle$ denote a nonequilibrium expectation, (i.e., over trajectories $x(t)|_{t \in [0, t_f]}$ that start in equilibrium with $\lambda_i$ and evolve via some specified Langevin dynamics), while $\langle \cdot \rangle^\eq_{\lambda}$ denotes an expectation value over the equilibrium thermodynamic state corresponding to control parameters set to $\lambda$. 

The probability distribution over the system's microstates at equilibrium with respect to control parameters $\lambda$ is given by the canonical ensemble,
\begin{equation}
    \peq = \text{exp}\{-\beta [H_\lambda(x) - F(\lambda)]\} = \frac{e^{-\beta H_\lambda(x)}}{Z(\lambda)}, \label{eq:rho-eq}
\end{equation}
where $\beta$ is the inverse temperature, and $F$($\lambda$) = $-\beta^{-1}$ ln[$\int$exp$\{-\beta H_\lambda(x') \}  \dx'$] is the free energy of the equilibrium state determined by $H_\lambda(x)$, which can also be written using the partition function $Z(\lambda) = \int\exp\{-\beta H_\lambda(x') \}  \dx' $.

In general, additional factors beyond energy such as the friction and noise distribution are required to determine the dynamics of the system, which will be fixed when specific examples are considered. We assume the time-evolution of the nonequilibrium probability density $\rho_t$ is governed by the following deterministic partial differential equation:
\begin{equation}
    \frac{\partial \rho_t}{\partial t} = \mathcal{L}_{\lambda} [ \rho_t],\label{eq:fokker-planck-eq}
\end{equation}
where in the continuous case $\mathcal{L}_\lambda$ is the Fokker-Planck operator. (For the discrete case, this would be the master equation and transition rate matrix.)

Here, we focus on passive, detailed-balance preserving systems, and so we will assume that for any $\lambda$ defining the potential energy function, the equilibrium distribution is the unique zero-mode of the $\mathcal{L}_\lambda$ operator 
\begin{equation}
  \mathcal{L}_{\lambda}[\rho^\mathrm{eq}_\lambda] = 0, \label{eq:zero-mode}
\end{equation}
and under the dynamics $\partial_t \rho_t = \mathcal{L}_\lambda [\rho_t]$ for a fixed $\lambda$, any arbitrary initial distribution $\rho_0$ will relax to the equilibrium distribution, $\lim_{t \rightarrow \infty} \rho_t = \rho^\mathrm{eq}_\lambda$. Additionally, we will only consider isothermal protocols, with a fixed friction coefficient in the continuous case. Because of this, from here 
on out we set inverse temperature $\beta = 1$. 

Splitting the nonequilibrium probability density $\rho_t(x)$ at time $t$ into the equilibrium density $\peq$ [fixed by the control parameters $\lambda(t)$] and nonequilibrium deviation $\delta \rho_t(x)$ at time $t$,
\begin{equation}
    \label{eq:equi_nonequi}
    \rho_t(x) = \peq(x) + \delta \rho_t(x),
\end{equation}
where, since $\peq(x)$ is the zero-mode of $\FokkerPlanck$ [Eq.~\eqref{eq:zero-mode}], we can rewrite the time evolution as
\begin{equation}
    \frac{\partial \rho_t}{\partial t} = \mathcal{L}_{\lambda} [\delta \rho_t].  \label{eq:fokker-planck-eq-zeromode-removed}
\end{equation}

Using $\delta \rho_t$, we can distinguish average excess work from average total work. Inserting Eq.~\eqref{eq:equi_nonequi} into Eq.~\eqref{eq:total_work} we obtain
\begin{equation}
    \langle W \rangle = \langle W_{\text{rev}} \rangle + \langle W_{\text{ex}} \rangle,
\end{equation}
where the two terms are
\begin{gather}
    \langle W_\mathrm{rev} \rangle = \int^{t_f}_0 \dt \dot{\lambda}^a \int  \d x \frac{\partial H_\lambda(x)}{\partial \lambda^a} \rho_\lambda^\eq(x) , \\
    \label{eq:excess_work}
    \langle W_\mathrm{ex} \rangle =  \int^{t_f}_0 \dt \dot{\lambda}^a \int  \d x \frac{\partial H_\lambda(x)}{\partial \lambda^a} \delta \rho_t(x).
\end{gather}
Here, $\langle W_\mathrm{rev} \rangle$ is the reversible work, which is equal to the free energy difference $\Delta F = F(\lambda_f) - F(\lambda_i)$ between the equilibrium states defined by the initial and final control parameters $\lambda_i$ and $\lambda_f$, which is independent of the protocol between the initial and final points. $\langle W_\mathrm{ex} \rangle$ is the excess work, which is dependent on the shape and speed of the protocol. 

\begin{figure}
    \centering
    \includegraphics[width=0.85\linewidth]{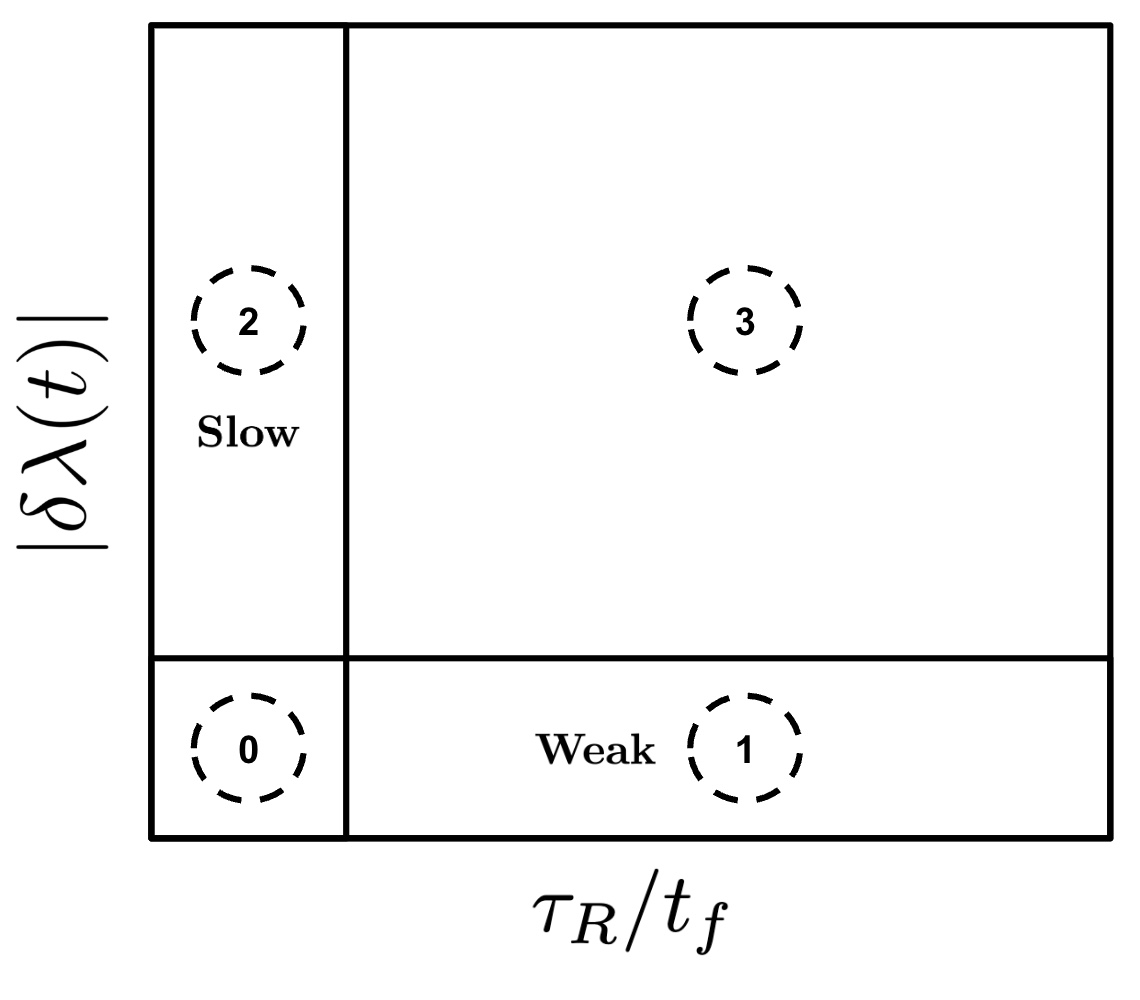}
    \caption{\textit{Approximation regimes in optimal stochastic thermodynamics.} Following \cite{dibyenduperscom,bonanca2018}, we can parameterize the control protocol as $|\delta\lambda(t)| = |\lambda(t) - \lambda_i|$, and $\tau_R$ represents an appropriate relaxation time for the system. Region 0 corresponds to weak and slow driving, Region 1 corresponds to weak driving, Region 2 corresponds to slow driving, and Region 3 corresponds to neither weak nor slow driving, where protocols push the system further from equilibrium.}
    \label{fig:approx_regimes}
\end{figure}

Given that $\langle W_\mathrm{rev}\rangle = \Delta F$ is simply a function of the initial and final equilibrium states, our task becomes minimizing Eq.~\eqref{eq:excess_work}. The problem of minimizing average excess work can be broken into roughly four regimes where different approximations apply, as illustrated in Fig.~\ref{fig:approx_regimes}: corresponding to slow driving ($t_f$ is ``large'', Region 2), weak driving (the two equilibrium distributions specified by $\lambda_i$ and $\lambda(t)$ are nearby in a suitable sense, Region 1), simultaneously slow and weak driving (Region 0), and driving which is neither slow nor weak (Region 3). 

In the past, linear-response-based approaches have been developed to find the optimal protocol minimizing the excess work. For the slow-driving limit, \cite{sivak2012} developed a Riemannian-geometric based approach to calculate optimal protocols minimizing Eq.~\eqref{eq:total_work}, showing that the excess work can be approximated by
\begin{equation}
    \langle W_\mathrm{ex} \rangle \approx \int_0^{t_f} \dot{\lambda}^a \dot{\lambda}^b g_{a b}(\lambda(t)) \, \dt, \label{eq:SV-cost}
\end{equation}
where $g_{a b}(\lambda)$ is the friction tensor that may be readily computed using equilibrium time-correlation functions as we detail below. This friction tensor can be interpreted as a Riemannian metric on the space of control parameters $\lambda \in \mathcal{M}$, and under this interpretation, Eq.~\eqref{eq:SV-cost} is squared path divergence for the protocol $\Lambda$ in the Riemannian geometry induced by $g_{a b}(\lambda)$. Thus, in this appropriate slow-driving limit, the optimal protocol is a geodesic connecting $\lambda_i$ and $\lambda_f$, yielding the excess cost as a squared thermodynamic length \cite{crooks2007measuring} in this geometry. 

Alternatively, \cite{bonanca2018, naze2022optimal, naze2024analytical} 
considered the weak driving limit, where $\delta \lambda(t) / \lambda_i $ is suitably small (or that $\lambda(t) \approx \lambda_i$ throughout the protocol), showing that the excess work may instead be expressed as 
\begin{equation}
    \langle W_\mathrm{ex} \rangle \approx \frac{1}{2} \int_0^{t_f} \int_0^{t_f} \dot{\lambda}^a(t) \, \Psi^{(1)}_{a b}(\lambda_i; |t - t_1| ) \, \dot{\lambda^b}(t_1) \, \dt \, \dt_1, \label{eq:weak-driving-cost-fnc}
\end{equation}
where $\Psi^{(1)}_{a b}(\lambda_i; \tau)$ is the so-called ``relaxation function" at time-lag $\tau$ defined for $\lambda_i$, that may either be computed similarly using an average over the initial equilibrium state, or constructed phenomenologically. 

These approximations rely on low control parameter velocity in the slow driving case (Regions 0 and 2 of Fig.~\ref{fig:approx_regimes}), and low variations from the initial control parameters in the weak driving case (Regions 0 and 1). As we will see, each of these results can be derived from the first order of a Volterra series. 

A natural idea for attempting to extend approximations from Regions 1 or 2 into Region 3 (neither slow nor weak, further from equilibrium) is to include higher-order terms of the series approximation on which they depend. Both the weak and slow driving approximations rely on linear response: to extend them, we will consider including higher order response. In the following section, we will introduce the Volterra series, discuss its application to optimal stochastic thermodynamics, and show how these approximations naturally arise from its first order.

\section{The Volterra Expansion of $\boldsymbol{\delta \rho_t}$}

We can begin by suitably rewriting Eq.~\eqref{eq:excess_work}, then considering the response of the equilibrium correction $\delta \rho_t$ to changes in the underlying Hamiltonian. Defining the conjugate force with respect to control parameter $\lambda^a$ as 
\begin{equation}
    f_a(x) := - \frac{\partial H_\lambda(x)}{\partial \lambda^a} \label{eq:conjugate-force}
\end{equation}
we can write Eq.~\eqref{eq:excess_work} as
\begin{equation}
    \langle W_\text{ex} \rangle = -\int^{t_f}_0 \dt \dot{\lambda}^a \int  \dx f_a (x) \neqcorrection(x). 
\end{equation}
Equivalently, because the normalization $\int \rho_t(x) \dx = 1$ and $\int \rhoeq_\lambda (x) \dx = 1$ implies that $\int \delta \rho_t(x) \dx = 0$, the excess work can be rewritten as

\begin{equation}
    \label{eq:excess-work-general}
    \langle W_\text{ex} \rangle = -\int^{t_f}_0 \dt \dot{\lambda}^a \int  \d x \, \delta f_a (x) \,\neqcorrection(x),
\end{equation}
where
\begin{equation}
   \label{eq:conjugate_force}
    \delta f_a(x) = f_a(x) - \langle f_a \rangle^\mathrm{eq}_\lambda 
\end{equation}
is the excess conjugate force. Note that one can relate the excess conjugate force and the equilibrium probability density via the identity
\begin{equation}
    \label{eq:conjugate_identity}
    \delta f_a(x) = \frac{\partial \ln \rho^{eq}_{\lambda}(x)}{\partial \lambda^a}.
\end{equation}

Borrowing from control theory, 
$\neqcorrection$ can be systematically approximated using a Volterra series \cite{boyd1984analytical, BROCKETT1976167, wiener1942response, FRISTON20031273, lesiak1977}. The Volterra series considers the response of a dynamical system to a series of inputs. In finite-time thermodynamics, the system is driven between equilibria, where at each point it is pushed out of equilibrium to a variable extent by changes in $\lambda(t)$. This suggests a natural identification of the vector of control parameter velocities $\dot{\lambda}$ as the input variables to the dynamical system, with the deviation from equilibrium being the output (as $\dot{\lambda} = 0$ leads to $\delta \rho_t = 0$). For this choice, impulses in $\dot{\lambda}$ correspond to steps in the protocol $\Lambda = \lambda(t) |_{t\in[0,t_f]}$. 

Given this identification, the Volterra series approximation for $\delta \rho_t$ is given by 
\begin{widetext}
\begin{equation}
    \label{eq:full_volterra}
    \begin{aligned}
        \delta \rho_t(x) \approx 
        \sum^N_{n=1} \int^t_{-\infty} ... \int^t_{-\infty} \bigg\{ \dot{\lambda}^{i_1}(t_1) \cdot ... \cdot \dot{\lambda}^{i_n}(t_n) \cdot 
       h^{(n)}_{i_1 ... i_n}(\lambda(t); t-t_1, ... , t-t_n) 
      \bigg\} \prod^n_{j=1} \dt_j . 
    \end{aligned}
\end{equation}
\end{widetext}
Here $h^{(n)}_{i_1... i_n}(\lambda(t); t-t_1, ... , t-t_n)$ is the $n^{\text{th}}$-order Volterra kernel defined 
for the system fixed by control parameters $\lambda(t)$, which can be evaluated by considering $n$ impulses to the system at the timelags $t - t_1, ..., t-t_n$, as we shall explain below. (As a reminder, under Einstein summation notation each $i_1$, $i_2$, ..., sums from $1$ to $k$, the dimensionality of the control space manifold $\mathcal{M}$.) Here, each Volterra kernel will be a function of position $h^{(n)}_{i_1...i_n}(\lambda(t);...)(x)$ representing the response at each point $x$; however, from here onward the $x$-dependence will be suppressed for clarity. For lengthy expressions, the $x$-dependence of conjugate forces $\delta f(x)$ or probability densities $\rho(x)$ will be similarly suppressed.

\subsection{First Order - Linear Response}

\begin{figure*}
    \centering
    \includegraphics[width=0.8\linewidth]{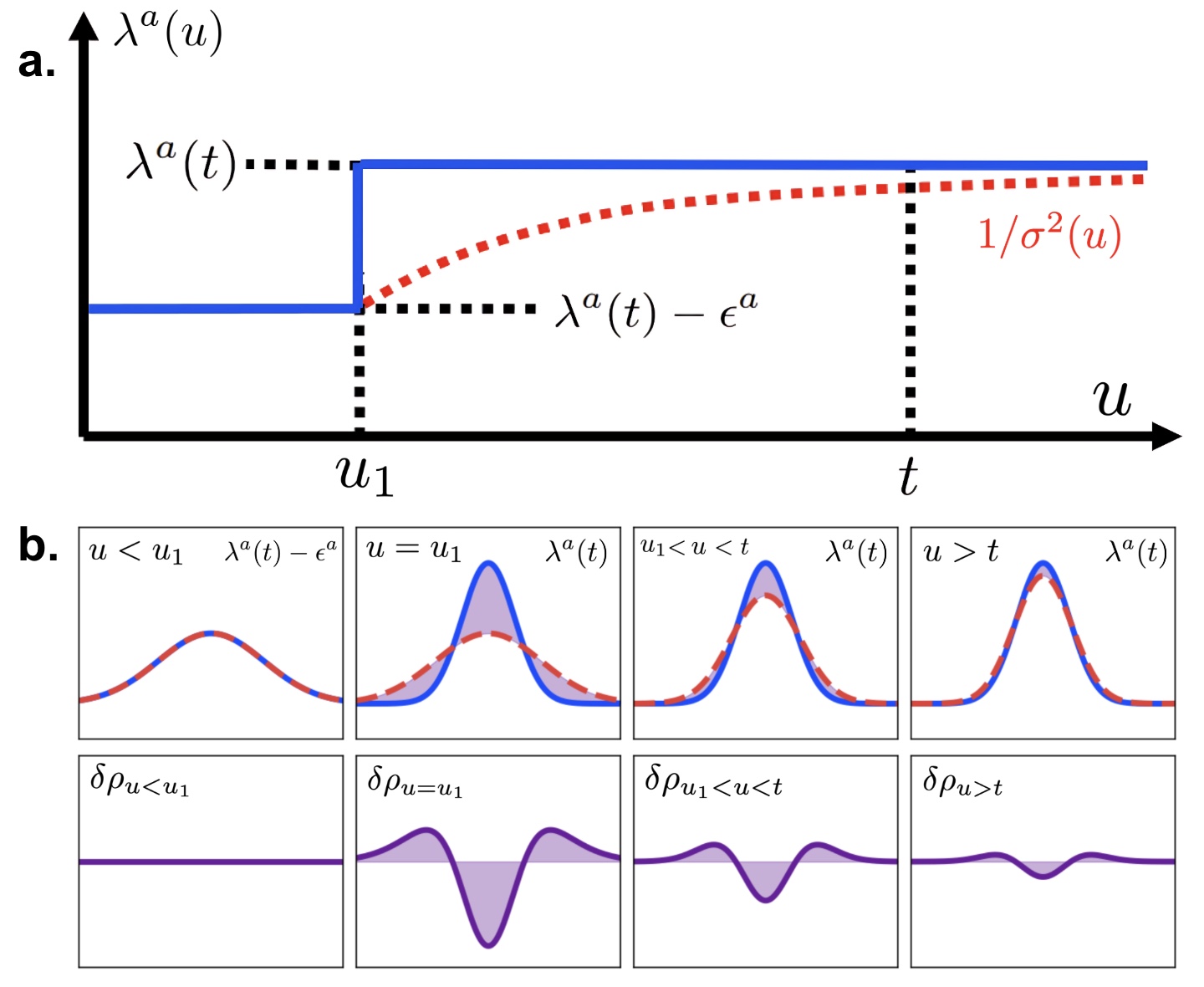}
    \caption{\textit{Derivation of the linear response function.} \textbf{a.} The control parameter $\lambda^a$ moves from $\lambda^a(t) - \epsilon^a$ to $\lambda^a(t)$ at $u_1$. If we assume $\lambda^a(u) = \alpha(u)$, where $\alpha(u)$ is the stiffness of a simple harmonic oscillator, the inverse variance of the nonequilibrium distribution, $1/\sigma^2(u < u_1)$, is equal to $\alpha(u < u_1)$ at equilibrium. Once the control parameter changes, $\delta \rho \neq 0$ and $1/\sigma^2(u) \neq \alpha(u)$. After this, over time $\delta \rho$ relaxes back to zero and $1/\sigma^2(u)$ relaxes to the new value, $\alpha(t)$. 
    This picture is exact for the simple harmonic oscillator, but in general, this serves as a cartoon as the nonequilibrium density will not correspond to an equilibrium density for the same system for any control parameter value, requiring more than a single parameter to characterize their difference.
    \textbf{b.} 
    The evolution of the nonequilibrium density $\rho_u$ (dotted red line) is shown along with the equilibrium density $\rhoeq_{\lambda(u)}$ (solid blue line) corresponding to the current control parameter value (as specified in the upper right corner of the upper panels), with their difference indicated by the purple shaded area. 
    In the bottom panels, assuming $\lambda^a(u) = \alpha(u)$, the stiffness of a simple harmonic oscillator, the equilibrium distribution defined by the potential abruptly changes from $\rhoeq_{\lambda^a(t)-\epsilon^a}$ to $\rhoeq_{\lambda^a(t)}$. Initially $\delta \rho_u = 0$. After $u = u_1$, $\delta \rho_u$ decays back to zero as $\rho_u \rightarrow \rhoeq_{\lambda(t)}$.}
    \label{fig:linear-response}
\end{figure*}

In this section we show how both thermodynamic geometry and weak-driving linear-response approximations [Eqs.~\eqref{eq:SV-cost} and \eqref{eq:weak-driving-cost-fnc}] can be derived from this Volterra series expansion truncated at $N = 1$. The nonequilibrium correction given by first-order response is defined as 
\begin{equation}
    \label{eq:linear-response-correction}
    \delta \rho^{(1)}_t = \int^t_{-\infty} h^{(1)}_{a}(\lambda(t); t - t_1) \dot{\lambda}^{a}(t_1) \dt_1.
\end{equation}
The first-order Volterra kernel (equivalently called first-order or linear response function) $h^{(1)}$ can be evaluated by considering the system's first-order response to a jump perturbation. This jump perturbation in the control parameter $\lambda^a$ can be expressed as
\begin{equation}
    \label{eq:velocity-perturb-linear}
    \dot{\lambda}^a(u) = \epsilon^a \delta(u - u_1),
\end{equation}
where $u$ is a generic time along the jump protocol, and $\epsilon$ is a small-component control vector, corresponding to a single step which moves the parameter $\lambda^a$ from $\lambda^a(t) - \epsilon^a$ to $\lambda^a(t)$ at time $u_1$,  (where $\lambda(t)$ is the vector of control parameters at the time $u = t$, where $t > u_1$). We can find the first-order response function by evaluating the response of $\delta \rho_u$ to the step to first order in $\epsilon^a$, assuming the system starts in equilibrium with the perturbed control state $\lambda(t) - \epsilon$ (see Fig.~\ref{fig:linear-response}). We will refer to the equilibrium state of the perturbed control state as $\rho^{\eq}_{\lambda(t) - \epsilon} = \rho^{\eq}_1$, and the unperturbed equilibrium state as $\rho^{\eq}_{\lambda(t)} =\rho^{\eq}_0$. 

Following Eq.~\eqref{eq:equi_nonequi}, the nonequilibrium correction just after the jump is 
\begin{equation}
    \label{eq:noneq_corr_after_1_jump}
    \delta \rho_{u_1}(x) = \rhoeq_{1}(x) - \rhoeq_{0}(x).
\end{equation}
Linearizing with respect to $\epsilon$, and applying the identity $\partial_{\lambda^a} (\rhoeq_\lambda) = \rhoeq_\lambda \, ( \partial_{\lambda^a} \ln \rhoeq_\lambda )
= \peq \delta f_a$ (using Eq.~\eqref{eq:conjugate_identity}, with $\partial_{\lambda^a} = \partial / \partial \lambda^a$), we have 
\begin{equation}
    \label{eq:linear_diff}
    \rhoeq_{1} = \rhoeq_{0}[ 1 - \epsilon^a \delta f_a ] + \mathcal{O}(\epsilon^2).
\end{equation}
To approximate the non-equilibrium correction at $u=t$, $\delta \rho_t$, we insert Eq.~\eqref{eq:linear_diff} to Eq.~\eqref{eq:noneq_corr_after_1_jump} and apply the unperturbed time evolution operator exp$(\mathcal{L}_{0} (t-u_1))$ to obtain $\delta \rho_{t}$ up to linear order  (where we use the notation $\mathcal{L}_{\lambda(t)} = \mathcal{L}_0$):
\begin{equation}
    \label{eq:delta_rho_1}
    \begin{aligned}
        \delta \rho_{t} \approx \delta \rho^{(1)}_t =  \underbrace{-\epsilon^a \text{exp}(\mathcal{L}_0 (t - u_1)) [\rhoeq_{0} \delta f_a]}_{\epsilon^ah^{(1)}_a(\lambda(t); t - u_1) },
    \end{aligned}
\end{equation}
where the matching is obtained by inserting Eq.~\eqref{eq:velocity-perturb-linear} into Eq.~\eqref{eq:linear-response-correction}. This shows that the first-order response function for $\delta \rho_t$ is 
\begin{equation}
    \label{eq:linear-response-function}
    h^{(1)}_a(\lambda(t); \tau) = -\exp(\mathcal{L}_{\lambda(t)} \tau) [\rho^\text{eq}_{\lambda(t)} \delta f_a].
\end{equation}

With this, we can write out the excess work assuming linear response, $\langle W_\text{ex}^{(1)} \rangle$, by replacing $\delta \rho_t$ in Eq.~\eqref{eq:excess-work-general} with $\delta \rho_t^{(1)}$ with the above approximations 
\begin{widetext}
\begin{equation}
    \label{eq:Wex1}
    \langle W_\text{ex}^{(1)} \rangle = \int^{t_f}_0 \dt \dot{\lambda}^a(t) 
    \int^{t}_{-\infty} \dt_1 \dot{\lambda}^b(t_1) 
    \underbrace{\bigg\{ \int  \d x \, \delta f_a(x)  \, \text{exp}(\mathcal{L}_{\lambda(t)} (t-t_1)) [\delta f_b(x) \rhoeq_{\lambda(t)}(x)] \bigg\}}_ {\Psi_{ab}^{(1)} = \,-\int \, dx \,\delta f_a(x) \, h^{(1)}_b(x)}
\end{equation}
\end{widetext} 
This formula can be matched to previous results where linear response is introduced by observing that the relaxation function (as defined by Eq.~\eqref{eq:Wex1}) is the excess conjugate force correlation function
\begin{equation} 
    \label{eq:linear-relaxation}
    \begin{aligned}
        \Psi_{a b}^{(1)}(\lambda(t); \tau) = & \int  \d x \delta f_a(x)  e^{\mathcal{L}_{\lambda(t)}\tau} [ \delta f_b(x) \rhoeq_{\lambda(t)}(x)] \\ & = \langle \delta f_a(0) \delta f_b(\tau) \rangle^\eq_{\lambda(t)}.
    \end{aligned}
\end{equation}

From this expression, we can derive thermodynamic geometry and the weak-driving approximations.

\subsubsection{Low Acceleration Limit (Thermodynamic Geometry)}

With an additional assumption, the results from thermodynamic geometry can be derived from the first-order term of the Volterra series of $\delta \rho_t$. By Taylor-expanding the control parameter velocity around its value at $t$ and assuming the acceleration of control parameters is small, we arrive at the thermodynamic geometry results used in \cite{sivak2012, Wadia_2022, sawchuk2024dynamical, zulkowski2012geometry, rotskoff2015optimal, rotskoff2017geometric, zulkowski2015optimal, frim2022geometric, frim2022optimal}. This additional step means thermodynamic geometry assumes slow driving by taking the first term of the Volterra series, and low acceleration driving by using the expansion
\begin{equation}
    \dot{\lambda}(t_1) = \dot{\lambda}(t) + \mathcal{O}(\ddot{\lambda}(t)), 
\end{equation}
which gives us a time-local approximation for the excess power $d\langle W_{ex}\rangle /dt$. All time-local approximations will be marked with the superscript $\text{tl}$. Thermodynamic geometry assumes $\langle W_{ex}^{(1)}\rangle \approx \langle W_{ex}^{(1),\text{tl}}\rangle$ by assuming $\dot \lambda(t_1) \approx \dot \lambda(t)$, where we have 
\begin{equation}
    \langle W_\text{ex}^{(1),\text{tl}} \rangle = \int^{t_f}_0 \dt \dot{\lambda}^a(t)\dot{\lambda}^b(t) \int^t_{-\infty} \dt_1\Psi_{a b}^{(1)}(\lambda(t); t-t_1).
\end{equation}

Changing variables to $\tau_1 = t - t_1$, we can rewrite the integral on the right of the above expression as the friction tensor,
\begin{equation}
    \label{eq:first-order-correction}
    \begin{aligned}
        \zeta^{(1)}_{a b}(\lambda(t)) & = \int^{\infty}_0 d\tau \Psi_{a b}^{(1)}(\lambda(t); \tau ) \\
        & = \int^{\infty}_0 d\tau \langle \delta f_a(0) \delta f_b(\tau) \rangle^\eq_{\lambda(t)}
    \end{aligned}
\end{equation}
\begin{equation}
    \label{eq:smooth}
    \langle W_\text{ex}^{(1),\text{tl}} \rangle = \int^{t_f}_0 \dt \dot{\lambda}^a(t)\dot{\lambda}^b(t) \zeta^{(1)}_{a b}(\lambda(t)),
\end{equation}
which is the standard expression used to define a metric on the space of thermodynamic states \cite{sivak2012, zulkowski2012geometry, rotskoff2015optimal, rotskoff2017geometric, zulkowski2015optimal, frim2022geometric, frim2022optimal}. Geodesics of this metric provide optimal protocols assuming both low control parameter acceleration and low control parameter velocity, offering a computationally inexpensive way to calculate optimal protocols, but its near-optimality in performance is only guaranteed if these  assumptions are satisfied \cite{sivak2012, zulkowski2012geometry}. (See, e.g., \cite{zhong2022limited} where the geodesic protocol significantly under-performs outside of the slow-driving regime.) 

Recent work on thermodynamic geometry has further grounded this result by deriving it from a perturbative time-local series \cite{Wadia_2022} and using the Drazin inverse \cite{sawchuk2024dynamical}. We can recover the time-local results presented in \cite{Wadia_2022} and \cite{sawchuk2024dynamical} from Eq.~\eqref{eq:Wex1} by writing the time evolution operator in terms of $\text{Pr}_{\lambda(t)}(x,t|y,t_1)$, the probability of finding the system in state $x$ at time $t$ given the system was initialized in state $y$ at time $t_1$, and that the control state of the system is at $\lambda(t)$ between $t_1$ and $t$:
\begin{equation}
    e^{\mathcal{L}_{\lambda(t)}(t-t_1)}[g(\cdot)](x) = \int dy \, \text{Pr}_{\lambda(t)}(x,t|y,t_1) g(y).
\end{equation}
From this, we can write the Drazin inverse as \cite{sawchuk2024dynamical}: 
\begin{equation}
    \label{eq:drazin}
    \begin{aligned}
         \mathcal{L}_{\lambda(t)}^{-1}[g(\cdot)] = - \int_0^{\infty} & d \tau \bigg[
         \int dy \,  \text{Pr}_{\lambda(t)}(x,\tau|y,0) g(y)\\ & - \rhoeq_{\lambda(t)}(x) \int dy \, g(y) \bigg].
    \end{aligned}
\end{equation}
To write the excess work in terms of the Drazin inverse, the low acceleration approximation must be used to fix $\dot \lambda(t_1)$ to $\dot \lambda(t)$. This allows $\dot \lambda (t)$ to be moved into the integral, meaning $\dot \lambda^a(t) \partial_a \rhoeq_{\lambda(t)} = \partial_t \rhoeq_{\lambda(t)}$. Since $\int dy \, \partial_t \rhoeq_{\lambda(t)} = 0$, the second term in Eq~\eqref{eq:drazin} vanishes, implying that the first order cost function is
\begin{equation}
    \langle W^{(1),\text{tl}}_\text{ex} \rangle \approx -\int^{t_f}_0 \d t \dot \lambda^a(t) \int dx\, \delta f_a(x) \mathcal{L}^{-1}_{\lambda(t)}[\partial_t \rhoeq_{\lambda(t)}](x)
\end{equation}

The extension of this approximation to higher orders is discussed in Sec.~\ref{sec:higher_time_local_order}, and derived in detail in App.~\ref{app:drazin_equivalence}. Additionally, in App.~\ref{app:harmonic}, we consider higher orders of the time-local perturbative expansion for an overdamped moving harmonic trap, observing that the inclusion of higher orders can lead to unphysical negative excess work if evaluated outside its slow-driving range of applicability. Additionally, true optimal protocols often include finite jumps \cite{seifert2007, zhong2024, aurell2011optimal, Aurell2012BoundaryLayers}, which lead the first term in the time-local series [Eq.~\eqref{eq:smooth}] to diverge, in contrast to the time-nonlocal approximation given by Eq.~\eqref{eq:Wex1}. 

\subsubsection{Weak Driving Limit}

The weak driving result from~\cite{bonanca2018} can be derived from a simple change to the first term in the Volterra series Eq.~\eqref{eq:Wex1}. $\Psi_{a b}^{(1)}(\lambda(t); t-t_1)$ is approximated by an \textit{ansatz} defined by the initial state of the system. In the overdamped limit this is given by 
\begin{equation}
    \label{eq:weak_approx}
    \Psi_{a b}^{(1)}(\lambda(t); \tau_1) \approx \Psi_{a b}^{(1)}(\lambda(0); 0) \cdot \exp(-s |\tau_1|),
\end{equation}
where $s$ is a phenomenological parameter~\cite{bonanca2018}. This is computationally advantageous since the left hand side of Eq.~\eqref{eq:weak_approx} is dependent on both $t$ and $|t-t_1|$, whereas the right hand side depends only on $|t-t_1|$. Expanding the protocol with a set of basis functions, $\langle W_\text{ex}^{(1)} \rangle$ can be rewritten in terms of the protocol's basis coefficients, which is possible because the \textit{ansatz} is independent of $\lambda(t)$. In this case $\langle W_\text{ex}^{(1)} \rangle$ can be efficiently optimized since only the basis coefficients need to be varied.

The \textit{ansatz} setting the dependence on $\tau_1$ is further discussed in \cite{bonanca2014}. However, in this case we can consider the spectral properties of the Fokker-Planck operator to assess the validity of this \textit{ansatz}. In the weak driving limit we can simply say $\lambda(t) \approx \lambda(0)$, so that
\begin{equation}
    \Psi_{a b}^{(1)}(\lambda(t); \tau_1) \approx \Psi_{a b}^{(1)}(\lambda(0); \tau_1).
\end{equation}

Recalling Eq.~\eqref{eq:Wex1}, we can notice that the dependence on $\tau_1$ will be inherited from the exponential Fokker-Planck operator. If $\partial\rhoeq_\lambda(x) / \partial \lambda^b =  \delta f_b(x) \rhoeq_\lambda(x)$ is an eigenvector of the Fokker-Planck operator, the relaxation function will be a simple exponential with $-s$ equaling the corresponding eigenvalue. As we will see in Section III, this is true for an overdamped particle in a simple harmonic oscillator. However, if $\delta f_b(x) \rhoeq_\lambda(x)$ is not an eigenvector of $\FokkerPlanck$, this ansatz will be less appropriate.

It is interesting to note that \cite{naze2024analytical} found that for cost functions of the form Eq.~\eqref{eq:weak-driving-cost-fnc}, no matter the form of $\Psi_{ab}^{(1)}$, as long as it is a positive kernel that is unchanging over the protocol as given by the weak-driving assumption, exact solutions are always symmetric in time. This point is reflected in our numerical results in Section III.

\subsection{Second Order - Quadratic Response}

Next we turn to higher-order terms in the Volterra expansion for $\delta \rho_t$. These can be computed by considering the response to a sequence of steps rather than one, as in the linear response case. To clarify this, in this section we can explicitly evaluate the quadratic response. In the next section, we will extend these results to $N^{th}$ order. We can express the quadratic response approximation as
\begin{equation}
    \delta \rho_t \approx \delta \rho^{(1)}_t + \delta \rho^{(2)}_t,
\end{equation}
where $\delta \rho^{(1)}_t$ is given by Eq.~\eqref{eq:linear-response-correction}, and 
\begin{equation}
    \label{eq:quadratic-response-correction}
    \delta \rho^{(2)}_t = \int_{-\infty}^t \int_{-\infty}^t h_{a b} ^{(2)}(\lambda(t); t - t_1, t - t_2) \dot{\lambda}^a(t_1) \dot{\lambda}^b(t_2) \, \dt_1 \, \dt_2 .
\end{equation}
\begin{figure*}
    \centering
    \includegraphics[width=0.95\linewidth]{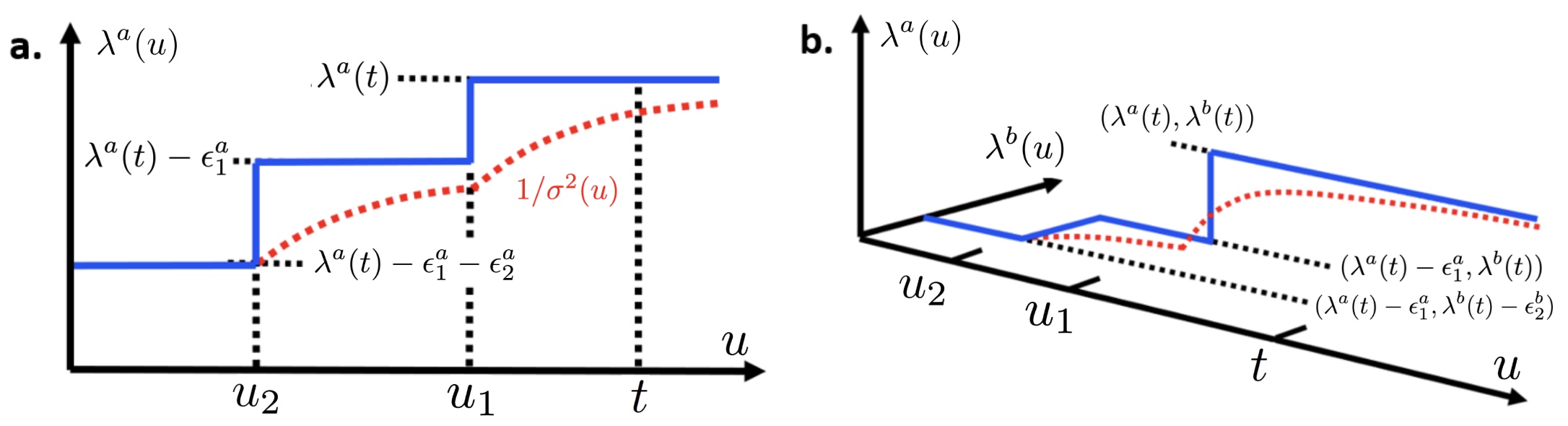}
    \caption{\textit{Derivation of the quadratic response function.} \textbf{a.} Response of the system (dotted red line) to two perturbations of a single control parameter $\lambda^a$ (solid blue line). For example, if the control parameter were the stiffness of a harmonic oscillator, the inverse variance of the nonequilibrium distribution would react to changes in this parameter.  Again, this picture serves as a cartoon for systems with densities not fully characterized by a single parameter. \textbf{b.} We also consider perturbations to each of two different control parameters, $\lambda^a$ and $\lambda^b$.}
    \label{fig:quadratic-response}
\end{figure*}

The quadratic response function can be evaluated by considering the second order response to a protocol in which the system starts in equilibrium with respect to two control perturbations, corresponding to the control state $\lambda(t) - \epsilon_1 - \epsilon_2$, which resolve through two jumps at $u_1$ and $u_2$ which take us to final control setting $\lambda_0 = \lambda(t)$. The velocity of this protocol can be written as
\begin{equation}
    \label{eq:quadratic_velocities}
    \dot{\lambda}^a(u) = \epsilon_1^a \delta(u - u_1)+ \epsilon_2^a \delta(u - u_2).
\end{equation}
Fig.~\ref{fig:quadratic-response} shows a cartoon of this process for two cases: when $\epsilon_1$ and $\epsilon_2$ perturb the same control parameter and when they perturb different control parameters. 

We will generalize our subscript shorthand from the previous section, such that for a control state which differs from $\lambda(t)$ by $N$ perturbations, $\lambda_N(t) = \lambda(t) - \sum_{n=1}^N\epsilon_n$, we will use the notation $\rhoeq_{\lambda_N} = \rhoeq_N$, $\mathcal{L}_{\lambda_N} = \mathcal{L}_N$, and $\int ( \cdot ) \rhoeq_{N}  \dx = \langle \cdot \rangle_N$. Similarly, we can write $\delta f_a^{(N)}$ as the excess conjugate force evaluated with respect to the equilibrium state corresponding to $\lambda_N$ (with the unadorned conjugate force being $\delta f_a = \delta f^{(0)}_a$).

Using these definitions, and applying the time evolution operator to the deviation from equilibrium at each step given by Eq.~\eqref{eq:noneq_corr_after_1_jump}, we can write the deviation from equilibrium at the time $t > u_1 > u_2$ as
\begin{equation}
    \label{eq:second_order_deviation}
    \begin{aligned}
        \delta \rho (t) = \,\,
        & e^{\mathcal{L}_0 (t-u_1)} \Big[ (\rhoeq_{1}- \rhoeq_{0})  + e^{\mathcal{L}_{1} (u_1 - u_2)} \big[ \rhoeq_{2} - \rhoeq_{1} \big] \Big].
    \end{aligned}
\end{equation}

This expression can be expanded to quadratic order in the perturbations to obtain the second-order response function. First, we can expand both $(\rhoeq_{1}- \rhoeq_{0})$ and $(\rhoeq_{2}- \rhoeq_{1})$ to second order. We can apply the identities
\begin{align}
    \partial_{\lambda^a} \partial_{\lambda^b} \rhoeq_0 &= \rhoeq_0(\delta f_a \delta f_b + \partial_{\lambda^b} (\delta f_a)), \\
    \partial_{\lambda^a} \partial_{\lambda^b} \rhoeq_1 &= \rhoeq_1(\delta f^{(1)}_a \delta f^{(1)}_b + \partial_{\lambda^b} (\delta f^{(1)}_a)),
\end{align}
to obtain
\begin{widetext}
\begin{equation}
    \label{eq:first_order_second_order}
    \begin{gathered}
        \delta \rho (t) = \,\, e^{\mathcal{L}_0 (t-u_1)} \bigg[ \rhoeq_0 \Big (- \epsilon_1^a \delta f_a + \frac{\epsilon_1^a\epsilon_1^b}{2} \Big[\delta f_a \delta f_b + \partial_{\lambda^b} (\delta f_a)\Big] + O(\epsilon_1^3)\Big) \\ + e^{\mathcal{L}_{1} (u_1 - u_2)}\bigg[ \rho^{eq}_1 \Big ( - \epsilon_2^a \delta f^{(1)}_a + \frac{\epsilon_2^a\epsilon_2^b}{2} \Big[\delta f^{(1)}_a \delta f^{(1)}_b + \partial_{\lambda^b} (\delta f^{(1)}_a) \Big ) \Big] + O(\epsilon_2^3)\bigg] \bigg].
    \end{gathered}
\end{equation}
\end{widetext}

With the first line of Eq.~\eqref{eq:first_order_second_order} already in terms of the unperturbed control state $\lambda(t)$, and the second line entirely in terms of the singly perturbed control state $\lambda(t) - \epsilon_1$, we can obtain the entire expression up to second order in $\epsilon_1$ and $\epsilon_2$ by expanding the entire second line in terms of $\epsilon_1$. We  only need to take the expansion of the second line to first order in $\epsilon_1$, as all terms in the expression are already to first order in $\epsilon_2$. 

\begin{widetext}
\begin{equation}
    \label{eq:first_order_second_order_next}
    \begin{gathered}
        \delta \rho (t) = \,\, e^{\mathcal{L}_0 (t-u_1)} \bigg\{ \rhoeq_0 \Big (- \epsilon_1^a \delta f_a + \frac{\epsilon_1^a\epsilon_1^b}{2} \Big[\delta f_a \delta f_b + \partial_{\lambda^b} (\delta f_a)\Big] + O(\epsilon_1^3)\Big) \\ 
        + e^{\mathcal{L}_{0} (u_1 - u_2)}\bigg[ \rho^{eq}_0 \Big( - \epsilon_2^a \delta f_a + \frac{\epsilon_2^a\epsilon_2^b}{2} \Big[\delta f_a \delta f_b + \partial_{\lambda^b} (\delta f_a)\Big]  + O(\epsilon_2^3) \Big)\bigg]
        -\epsilon_1^b \frac{\partial}{\partial \lambda^b} \Big [ e^{\mathcal{L}_{0} (u_1 - u_2)}\Big[ \rho^{eq}_0\Big( - \epsilon_2^a \delta f_a + O(\epsilon_1\epsilon_2^2) \Big) \Big] \bigg] \bigg\}.
    \end{gathered}
\end{equation}
\end{widetext}
With the expression fully expanded to second order around $\lambda(t)$, we can insert Eq.~\eqref{eq:quadratic_velocities} to Eq.~\eqref{eq:quadratic-response-correction} to determine the quadratic response function. The first two terms again recover the linear response function, while the following four terms give us the quadratic response function. From this expression we see that the quadratic response terms are higher order in terms of perturbations away from equilibrium. For weak protocols, this confirms that the quadratic response terms will be small. 

Gathering first order terms in Eq.~\eqref{eq:first_order_second_order_next} as $\delta \rho^{(1)}_t$, and second-order terms as $\delta \rho^{(2)}_t$, we can match this expression to that obtained by inserting Eq.~\eqref{eq:quadratic_velocities} into Eq.~\eqref{eq:quadratic-response-correction}. This allows us to perform the following matching:
\begin{widetext}
\begin{equation}
    \label{eq:quadratic_matching}
    \begin{aligned}
        \delta \rho_t \approx \delta \rho^{(1)}_t + &\,\, \delta \rho^{(2)}_t =  \underbrace{\epsilon_1^a \bigg \{ -e^{\mathcal{L}_0 (t - u_1)} [\delta f_a \rhoeq_0 ] \bigg\}}_{\epsilon^a_1 h^{(1)}_a(\lambda(t);t-u_1)} + \underbrace{\epsilon_2^a \bigg\{ -e^{\mathcal{L}_0 (t - u_2)} [\delta f_a \rhoeq_0 ] \bigg\}}_{\epsilon_2^a h^{(1)}_a(\lambda(t);t-u_2)}   
        + \underbrace{\frac{\epsilon_1^a \epsilon_1^b}{2}   \bigg\{ e^{\mathcal{L}_0 (t - u_1)} \big[ \rhoeq_0 (\delta f_a \delta f_b +\partial_{\lambda^b}(\delta f_a) ) \big ] \bigg \}}_{\epsilon_1^a\epsilon_1^b h_{ab}^{(2)}(\lambda(t); t-u_1,t-u_1)}
        \\ & \,\,\, + \, \underbrace{\frac{\epsilon_2^a\epsilon_2^b}{2} \bigg \{ e^{\mathcal{L}_0 (t - u_2)} \big[ \rhoeq_0 (\delta f_a \delta f_b
         +\partial_{\lambda^b}(\delta f_a) ) \big ] \bigg \}}_{\epsilon_2^a \epsilon_2^b h_{ab}^{(2)}(\lambda(t);t-u_2,t-u_2)}
        + \underbrace{\epsilon_1^b \epsilon_2^a \bigg \{ e^{\mathcal{L}_0 (t - u_1)}\Big[\frac{\partial}{\partial \lambda^b} \Big [ e^{\mathcal{L}_{0} (u_1 - u_2)} \big [ \rho^{eq}_0 \delta f_a \big ] \Big] \Big] \bigg \}}_{\epsilon_1^b\epsilon_2^a h_{ba}^{(2)}(\lambda(t); t-u_1,t-u_2)
        + \epsilon_1^b \epsilon_2^a h_{ab}^{(2)}(\lambda(t); t-u_2,t-u_1)}.
    \end{aligned}
\end{equation}
\end{widetext}
With the change of variables $\tau_n = t-u_n$, the quadratic response function can be identified as
\begin{equation}
    \label{eq:quadratic-response-function}
    h_{a b} ^{(2)}(\lambda_0; \tau_1, \tau_2) = \frac{1}{2} \begin{cases}
        e^{\mathcal{L}_0 \tau_1} \partial_{\lambda^a} \Big[  e^{\mathcal{L}_0 (\tau_2 - \tau_1)} [\delta f_b \rhoeq_0 ] \Big ], \tau_1 < \tau_2,
        \\ e^{\mathcal{L}_0 \tau_2} \partial_{\lambda^b} \Big[  e^{\mathcal{L}_0 (\tau_1 - \tau_2)} [\delta f_a \rhoeq_0 ] \Big ], \tau_2 < \tau_1.
    \end{cases}
\end{equation}

This can be used to evaluate the next-order approximation for the excess work, which can be written as
\begin{equation}
    \langle W_\mathrm{ex} \rangle \approx \langle W_\text{ex}^{(1)} \rangle + \langle W_\text{ex}^{(2)} \rangle,
\end{equation}
where, $\langle W^{(1)}_\text{ex} \rangle$ is given by Eq.~\eqref{eq:Wex1}, and 
\begin{widetext}
\begin{equation}
\langle W_\text{ex}^{(2)} \rangle = \int^{t_f}_0 \dt \dot{\lambda}^a(t) \int^{t}_{-\infty} \dt_1 \dot{\lambda}^b(t_1)  \int_{-\infty}^{t} \dt_2 \dot{\lambda}^c(t_2) \Psi_{a b c}^{(2)}(\lambda(t); t-t_1, t - t_2),
\end{equation}
\end{widetext}
with the next-order relaxation function $\Psi^{(2)}_{abc}$ given by \begin{equation}
    \Psi^{(2)}_{abc} = -\int dx \, \delta f_a(x) h_{bc}^{(2)}(x) 
\end{equation}

Explicitly computing the quadratic term helps clarify exactly why linear response approaches apply only in Regions 0, 1, and 2 of Fig.~\ref{fig:approx_regimes}. Eq.~\eqref{eq:quadratic-response-correction} shows that higher orders of the Volterra Series are of higher order in the control parameter velocity $\dot \lambda$. Slower manipulation of the control parameters will allow the linear term to dominate. Eq.~\eqref{eq:quadratic_matching} shows that higher orders of the Volterra Series contribute only for higher orders of perturbation strength $\epsilon$ from $\dot{\lambda}$, meaning weaker manipulation of control parameters allow the linear term to dominate. 

\subsection{Higher Order Response}

The same step-response construction used above can give us a recursive expression for the $n^{th}$ order response function, where
\begin{equation}
    \label{eq:nth_order_correction}
    \delta \rho^{(n)}_t = \int_{-\infty}^t...\int_{-\infty}^t h^{(n)}_{i_1...i_n}(\lambda(t); ...,t-t_n) \prod_{k=1}^n \dot \lambda^{i_k}(t_k) \d t_k,
\end{equation}
and the deviation from equilibrium can be given by
\begin{equation}
    \delta \rho_t \approx \sum_{n=1}^N \delta \rho_t^{(n)}
\end{equation}
With the final control parameter again given by $\lambda(t) = \lambda_0$, we can consider $n$ small step perturbations at $u_1, ..., u_n$, starting with the system in equilibrium with respect to $\lambda_0 - \sum_{j=1}^n \epsilon_j$. The velocity corresponding to this control protocol is given by

\begin{equation}
    \dot \lambda^a(u) = \sum_{j=1}^n \epsilon_j^a \delta(u - u_j).
\end{equation}
From the previous calculation, we can observe that the response function can be obtained by the cross term involving each perturbation $\epsilon_1, \epsilon_2, ... , \epsilon_n$. The terms involving higher orders of individual perturbations (ex: $\epsilon_1^2$) are derivable from the cross term with a symmetry factor (the factor of $1/2$ above). After $n$ perturbations, the generalization of Eq.~\eqref{eq:second_order_deviation} for $t > u_1 > u_2 > ... > u_n$ would be
\begin{equation}
    \label{eq:Nth_order_deviation}
    \begin{aligned}
        \delta \rho (t) = \,\,
        & e^{\mathcal{L}_0 (t-u_1)} \Big[ (\rhoeq_{1}- \rhoeq_{0})  + e^{\mathcal{L}_{1} (u_1 - u_2)} \big[(\rhoeq_{2} - \rhoeq_{1}) \\ + & \,e^{\mathcal{L}_2 (u_2-u_3)} \big [ ... + e^{\mathcal{L}_{n-1} (u_{n-1} - u_n)}[\rhoeq_n - \rhoeq_{n-1}]\big] ... \big] \Big].
    \end{aligned}
\end{equation}
Expanding this expression as was done above, and picking out only the cross term which includes all $\epsilon_1, \epsilon_2, ... , \epsilon_n$, we can obtain the general expression for the $n^{th}$ order response. Including the $1/n!$ symmetry factor, and again rewriting with $\tau_j = t-u_j$, this is
\begin{widetext}
\begin{equation}
    \label{eq:Nth_order_response}
    h_{i_1...i_n} ^{(n)}(\lambda_0; \tau_1, ... , \tau_n) = \frac{(-1)^n}{n!} e^{\mathcal{L}_0 \tau_1} \bigg [ \frac{\partial}{\partial \lambda^{i_1}} \bigg [ e^{\mathcal{L}_0 (\tau_2 - \tau_1)} \bigg [ \frac{\partial}{\partial \lambda^{i_2}} \bigg [ \, ... \,\, e^{\mathcal{L}_0(\tau_n - \tau_{n-1})} \Big[ \delta f_{i_n} \rhoeq_0 \Big]  \, ...\, \bigg ] \bigg] \bigg] \bigg],
\end{equation}
\end{widetext}
for $0 < \tau_1 < \tau_2 < ... < \tau_n$, where the partial $\partial / \partial \lambda^{i_m}$ derivatives act on $\mathcal{L}_0$, $\rhoeq_0$, and $\delta f_{i_n}$.

For an approximation truncated at $N^{th}$ order response, the excess work can be approximated as
\begin{equation}
    \label{eq:Wex_tot}
    \langle W_\text{ex} \rangle \approx \sum_{n=1}^N \langle W^{(n)}_\text{ex} \rangle,
\end{equation}
where with the compact notation $\boldsymbol{\tau} = (\tau_1, ..., \tau_n)$ we can write 
\begin{equation}
    \label{eq:Wex_n}
    \langle W^{(n)}_\text{ex} \rangle = \int_{0}^{t_f} dt \dot \lambda^{i_0}(t) \int_{\mathbb{R}^n_+} d^n \boldsymbol{\tau} \bigg[ \prod_{r=1}^n \dot \lambda^{i_r}(t-\tau_r) \bigg] \Psi^{(n)}_{i_0i_1...i_n}(\boldsymbol{\tau}),
\end{equation}
in which
\begin{equation}
    \label{eq:nth_order_relaxation}
    \Psi^{(n)}_{i_0i_1...i_n}(\lambda(t), \boldsymbol{\tau}) = -\int dx \, \delta f_{i_0} h^{(n)}_{i_1...i_n}(\lambda(t), \boldsymbol{\tau}).
\end{equation}

\subsubsection{Higher-order time-local series}

\label{sec:higher_time_local_order}

The $N^{th}$ order Volterra response series for the average excess work can be related directly to the time-local series discussed elsewhere \cite{Wadia_2022, sawchuk2024dynamical}. We can approach this by expanding the parameter velocities in the $n^{th}$ order response term given by Eq.~\eqref{eq:nth_order_correction} around $t_k = t$ as
\begin{equation}
    \dot \lambda^{i_k}(t_k) = \sum_{m_k=0}^\infty \frac{(t_k-t)^{m_k}}{(m_k)!} \frac{d^{m_k}}{d t^{m_k}} \dot \lambda^{i_k}(t).
\end{equation}
The time-dependent expansion is organized in orders of $\tau_k = (t - t_k)$. As described in App.~\ref{app:drazin_equivalence}, this implies that multiple Volterra orders will simultaneously contribute to the same order in the time-local series, with $m^{th}$ term in the expansion around $t$ is given by
\begin{align}
\label{eq:mth_order_time_local}
\delta\rho_{t}^{(m),\text{tl}}(x)
={}&
\sum_{n=1}^{m}
\sum_{\substack{m_1,\ldots,m_n\geq0\\
n+m_1+\cdots+m_n=m}}
\frac{(-1)^{m-n}}
{m_1!\cdots m_n!}
\nonumber\\
\times & 
\left[
\int_{\mathbb R_+^n}d^n\tau\,
\prod_{k=1}^{n}\tau_k^{m_k}
h_{i_1\ldots i_n}^{(n)}
\bigl(x;\lambda(t),\tau_1,\ldots,\tau_n\bigr)
\right]
\nonumber\\
\times & 
\prod_{k=1}^{n}
\frac{d^{m_k}}{d t^{m_k}}
\dot{\lambda}^{i_k}(t).
\end{align}

For example, the next order time-local approximation is given by:
\begin{equation}
    \label{eq:second_order_time_local}
    \begin{aligned}
        \delta \rho^{(2),\text{tl}}_t & = - \ddot \lambda^i(t)\int_0^\infty d\tau \, \tau \, h_i^{(1)}(\lambda(t);\tau) \\
        & + \dot \lambda^i(t) \dot \lambda^j(t) \int^\infty_0 d\tau_1 \int^\infty_0 d\tau_2 \, h_{ij}^{(2)}(\lambda(t); \tau_1, \tau_2).
    \end{aligned}
\end{equation}
The first term is due to the second time-order term from the expansion of the linear response term, while the second is the first time-order term in the quadratic response term.

In App.~\ref{app:drazin_equivalence}, we show that Eq.~\eqref{eq:mth_order_time_local} can be identified by successive applications of the Drazin inverse
\begin{equation}
    \label{eq:recursive_drazin}
    \delta \rho^{(m),\text{tl}}_t = \bigg ( \mathcal{L}^{-1}_{\lambda(t)} \frac{\partial}{\partial t} \bigg)^m \rhoeq_{\lambda(t)}.
\end{equation}
This identifies the relationship between the higher-order response series and the time-local series introduced elsewhere \cite{Wadia_2022, sawchuk2024dynamical}.

Inserting the series $\delta \rho_t \approx \sum_{m=1}^M \delta \rho_t^{(m),\text{tl}}$ into Eq.~\eqref{eq:excess-work-general} we can recover a higher-order time-local approximation for the average excess work. Writing this following Eq.~\eqref{eq:Wex_n}, and keeping the relaxation functions defined as in Eq.~\eqref{eq:nth_order_relaxation}, this can be expressed as
\begin{equation}
    \langle W_\text{ex} \rangle \approx \sum_{m=1}^M \langle W_\text{ex}^{(m),\text{tl}} \rangle,
\end{equation}
where
\begin{equation}
    \label{eq:mth_time_local_work}
    \begin{aligned}
        \langle W_\text{ex}^{(m),\text{tl}} \rangle & = \int_0^{t_f} \dot \lambda^{i_0}(t) \sum_{n=1}^m\sum_{\substack{m_1,\ldots,m_n\geq0\\ n+m_1+\cdots+m_n=m}} \frac{(-1)^{m-n}}
{m_1!\cdots m_n!} \\
        \prod_{k=1}^{n}
\frac{d^{m_k}}{d t^{m_k}}
\dot{\lambda}^{i_k} & (t) \int_{\mathbb R_+^n}d^n\tau\,
\prod_{k=1}^{n}\tau_k^{m_k}
\Psi_{i_0i_1\ldots i_n}^{(n)}
\bigl(\lambda(t),\tau_1,\ldots,\tau_n\bigr).
    \end{aligned}
\end{equation}

\subsubsection{Higher-order weak-driving series}

The weak driving series can also be evaluated to higher orders. In this case, the assumption is simply that higher orders of the Volterra series are included, but that they are set by the initial response functions. This means Eq.~\eqref{eq:Wex_tot} and Eq.~\eqref{eq:Wex_n} can be applied with the response function replaced by
\begin{equation}
    \Psi^{(n)}_{i_0i_1...i_n}(\lambda(0), \boldsymbol{\tau}) = -\int dx \, \delta f_{i_0} h^{(n)}_{i_1...i_n}(\lambda(0), \boldsymbol{\tau}),
\end{equation}
meaning the excess work is evaluated using the response kernels fixed at $\lambda(0)$. While optimization of the simplified 
higher-order work approximation may be accurate 
for protocols that 
stay close to $\lambda(0)$, we leave studying the effects of this additional weak-driving simplification 
to future work. 

\section{Linear and Higher-Order Response of the Overdamped Harmonic Oscillator}

\label{sec:harmonic}

In this section, we evaluate these approximations for the overdamped harmonic oscillator. Additionally, in App.~\ref{app:discrete_state} we evaluate them for a discrete two state system. For overdamped Langevin dynamics, with the system's microstate $x(t) \in \mathbb{R}$ and Gaussian white noise of unit amplitude $\eta(t)$, the system's trajectory is given by 
\begin{equation}
    \label{eq:langevin}
    \gamma \dot{x}(t) = -\frac{\partial \potentialCommand(x)}{\partial x} + \sqrt{2 \gamma \beta^{-1}} \eta(t),
\end{equation}
where $\gamma$ is the damping coefficient and $V_\lambda$ is the potential energy defined by control parameters $\lambda$. Since we will only consider protocols with constant temperature and damping coefficient, we set $\gamma = 1$ and $\beta$ = 1. In the overdamped case the linear Fokker-Planck operator $\FokkerPlanck$ is given by
\begin{equation}
    \FokkerPlanck[\rho] = \frac{\partial}{\partial x} \bigg[ \bigg( \frac{\partial \potentialCommand}{\partial x} +  \frac{\partial}{\partial x}\bigg) \rho\bigg]. 
\end{equation}
\subsection{Evaluation of Relaxation Functions}
For the overdamped harmonic case, we can evaluate the relaxation functions given by Eq.~\eqref{eq:nth_order_relaxation}, and thereby approximate the excess work Eq.~\eqref{eq:Wex_tot}. We can approach this using an eigenfunction decomposition, writing these expressions in terms of left $\rho_{l,n}(x)$ and right $\rho_{r,n}(x)$ eigenfunctions of the Fokker-Planck operator. 

In part, we adopt the harmonic example and the two-state example in App.~\ref{app:discrete_state} due to their evolution operators having convenient spectral decompositions, which make the evaluation of relaxation functions analytically tractable. In more complex cases, correlation functions may instead need to be evaluated numerically from stochastic trajectories (ex: \cite{rotskoff2015optimal} presents an example at linear order). The $k^{th}$-order relaxation functions depend on $k$ independent time lags. Estimating these increasingly high-dimensional functions from stochastic trajectories would require a substantial amount of simulated trajectory data to obtain statistically reliable results. 

In the harmonic case the left and right eigenfunctions are given by:
\begin{equation}
    \mathcal{L}_\lambda \rho_{r,n}(x) = r_n \rho_{r,n}(x),
\end{equation}
\begin{equation}
    \label{eq:right-eigenvector}
    \FokkerPlanck^\dagger \rho_{l,n}(x) = r_n \rho_{l,n}(x),
\end{equation}
which can be shown to satisfy 
\begin{equation}
    \rho_{r,n}(x) = \rhoeq_{\lambda(t)} \rho_{l,n}(x),
\end{equation}
and form a complete biorthogonal system \cite{Wadia_2022}
\begin{gather}
    \label{eq:dirac}
    \delta(x - y) = \sum_n \rho_{r,n}(x) \rho_{l,n}(y), \\ 
    \label{eq:orthonormal}
    \int  \dx \, \rho_{r,n}(x) \rho_{l,m}(x) = \delta_{nm}.
\end{gather}

For a one-dimensional harmonic oscillator with variable stiffness and center location
\begin{equation}
    \label{eq:SHO_potential}
    V_{\lambda(t)} = \frac{\alpha(t) (x - \mu(t))^2}{2},
\end{equation}
where $\lambda(t) = \big(\alpha(t), \mu(t)\big)$, the left eigenfunctions are 
\begin{equation}
    \label{eq:left-eigenvector}
    \rho_{l,n} = \frac{1}{\sqrt{n!}} \, \text{He}_n(\sqrt{\alpha(t)}(x-\mu(t))),
\end{equation}
with the eigenvalues
\begin{equation}
    r_n = -\alpha(t) \cdot n \quad\mathrm{for}\quad n = 0, 1, ... 
\end{equation}
where He$_n$ represents the $n^\mathrm{th}$ probabilist's Hermite polynomial \cite{Wadia_2022}. We can evaluate the response functions by applying the action of the Fokker-Planck time evolution operator on the right eigenfunctions 
\begin{equation}
    \label{eq:fokker_planck_time_evo}
    \begin{aligned}
        e^{\FokkerPlanck \tau} \rho_{r,n} = & \sum^\infty_{m=0} e^{r_m \tau} \rho_{r,m}(x) \int dy \, \rho_{l,m}(y) \, \rho_{r, n}(y) \\
         = & \,\, e^{-\alpha(t) n \tau} \rho_{r, n},
    \end{aligned}
\end{equation}
where we have used Eq.~\eqref{eq:orthonormal}.

Since the eigenfunctions form a complete basis, we can evaluate the response functions by using an eigenfunction expansion. For the linear response function, Eq.~\eqref{eq:linear-response-function}, the time evolution operator acts on the conjugate force. In the moving and stiffening harmonic case, the conjugate forces Eq.~\eqref{eq:conjugate_force} are given by
\begin{equation}
    \label{eq:harmonic_forces}
    \delta f_\alpha(x) = -\frac{\rho_{l,2}(x)}{\alpha \sqrt{2}}
    \quad\mathrm{and}\quad \delta f_\mu(x) = \sqrt{\alpha} \, \rho_{l,1}(x).
\end{equation}

Inserting Eq.~\eqref{eq:harmonic_forces} into Eq.~\eqref{eq:linear-relaxation} and applying Eqs.~\eqref{eq:fokker_planck_time_evo} and \eqref{eq:orthonormal}, we obtain the linear relaxation functions:
\begin{equation}
    \boldsymbol{\Psi}^{(1)}(\lambda(t); \tau) = 
    \begin{pmatrix}
        \frac{1}{2 \alpha^2} e^{-2\alpha \, \tau} & 0 \\
        0 & \alpha \, e^{-\alpha \, \tau} 
    \end{pmatrix}.
\end{equation}

This agrees with the low acceleration results for the friction tensor obtained elsewhere \cite{Wadia_2022, sivak2012, zulkowski2012geometry} by applying Eq.~\eqref{eq:first-order-correction} to obtain
\begin{equation}
    \zeta^{(1)}(\lambda(t)) =
    \begin{pmatrix}
        \frac{1}{4 \alpha^3} & 0 \\
        0 & 1 
    \end{pmatrix}.
\end{equation}

We can similarly evaluate the higher order relaxation functions and corresponding higher order slow-driving approximations. The general expression for the $n^{th}$ order relaxation function, with $0 < \tau_1 < ... <  \tau_n$ is
\begin{widetext}    
\begin{equation}
    \label{eq:nth-relaxation-expanded}
    \begin{aligned}
        \Psi^{(n)}_{i_0i_1...i_n} (\lambda(t); \boldsymbol{\tau}) & = \frac{(-1)^{n+1}}{n!}\int dx \, \delta f_{i_0}(x) e^{\FokkerPlanck \tau_1} \frac{\partial}{\partial \lambda^{i_1}} \Big [ e^{\FokkerPlanck (\tau_{2} - \tau_1)} \frac{\partial}{\partial \lambda^{i_2}} \Big [ ... \Big [\frac{\partial}{\partial \lambda^{i_{n-1}}} \Big [ e^{\FokkerPlanck (\tau_n - \tau_{n-1})} \Big ( \delta f_{i_n}(x) \rho^{eq}_\lambda(x) \Big) \Big ] ...\Big ] \Big ].
    \end{aligned}
\end{equation}
\end{widetext}
This expression can be evaluated by expanding $\delta f_{i_n}(x) \peq(x)$ in terms of the eigenfunctions of $\FokkerPlanck$, applying the time evolution operator to the resulting expression following Eq.~\eqref{eq:fokker_planck_time_evo}, then taking its partial derivative with respect to $\lambda^{i_{n-1}}$. Following this the expression can again be expanded in terms of eigenfunctions, the time evolution operator applied, and the next derivative taken with respect to $\lambda^{i_{n-2}}$. While this process will generate an increasing number of terms, only some will survive as $\delta f_{i_{0}}$ can be expanded in terms of Eq.~\eqref{eq:harmonic_forces}. Applying the orthogonality relation Eq.~\eqref{eq:orthonormal} to the resulting expression eliminates many terms. This process can be applied to find the $n^{th}$ order relaxation function.

Explicitly evaluating the quadratic relaxation functions, we can find that for $\tau_1 < \tau_2$ these are 
\begin{equation}
    \label{eq:quad_alpha_relaxation}
    \boldsymbol{\Psi}_\alpha^{(2)}(\lambda(t); \tau_1, \tau_2) = 
    \begin{pmatrix}
        \frac{(1+\alpha(\tau_2 - \tau_1))}{2 \alpha^3} \cdot e^{-2\alpha \tau_2} & 0 \\
        0 & \frac{1}{2} \cdot e^{-\alpha (\tau_1 + \tau_2)}  
    \end{pmatrix},
\end{equation}
\begin{equation}
    \label{eq:quad_mu_relaxation}
    \boldsymbol{\Psi}_\mu^{(2)}(\lambda(t); \tau_1, \tau_2) = 
    \begin{pmatrix}
        0 & \frac{\alpha(\tau_2 - \tau_1)}{2} \cdot e^{-\alpha \tau_2} \\
        0 & 0 
    \end{pmatrix} .
\end{equation}
(Where $\Psi_i = [[\Psi_{i \alpha \alpha}, \Psi_{i \alpha \mu}]^T, [\Psi_{i \mu \alpha}, \Psi_{i \mu \mu}]^T] $ is the notation used here.). The expressions for further higher-order response functions are given in App.~\ref{app:relaxation_functions}.

\subsection{Relation to Time-local Approximation}
This quadratic-response expression can be compared to previous results from Blaber and Sivak \cite{blaber2020skewed}, who derived a higher order term in $\langle W_\text{ex}\rangle$ by applying linear response to the squared excess work $\frac{\mathrm{d}}{\dt} \langle W_\mathrm{ex}^2 \rangle$ and using Crooks' fluctuation theorem \cite{crooks1999entropy, crooks2000path}, and used the stiffening harmonic oscillator as an example. Their second order approximation gives:
\begin{equation}
    \begin{aligned}
        \langle W_\text{ex} \rangle & \approx \int_0^{t_f} dt \, \dot \lambda^{a}(t) \dot \lambda^{b}(t)\bigg [ \zeta^{(1)}_{ab} + \dot \lambda^{c}(t)\frac{1}{4}\zeta^{(2)}_{abc} \bigg ] \\ & = \int_0^{t_f} dt \, \bigg [ \frac{\dot \alpha^2(t)}{4 \alpha^3} + \frac{\dot \alpha^3(t)}{8 \alpha^5} \bigg ],
    \end{aligned}
\end{equation}
where the first term is given exactly by $\langle W_\text{ex}^{(1),\text{tl}} \rangle$, and the second line is specific to the stiffening harmonic oscillator. 

This is comparable to the time-local series from Sec.~\ref{sec:higher_time_local_order}. Evaluating Eq.~\eqref{eq:mth_time_local_work} explicitly for second order, we can compare the resulting expression to their results for the stiffening harmonic trap.
\begin{equation}
    \label{eq:2nd_time_local_work}
    \begin{aligned}
    \langle W_\text{ex}^{(2),\text{tl}}\rangle &= \int_0^{t_f}dt \, \dot \alpha(t) \bigg [ - \ddot \alpha(t) \int_0^{\infty} d\tau \, \tau \, \Psi_{\alpha \alpha}^{(1)}(\tau)\\
    & + \dot \alpha (t) \dot \alpha(t) \int_0^{\infty} d\tau_1 \int_0^{\infty} d\tau_2 \,\Psi^{(2)}_{\alpha \alpha \alpha}(\tau_1,\tau_2) \bigg ].\\
    & = \int_0^{tf} dt \bigg [ - \frac{\dot \alpha(t) \ddot \alpha(t)}{8 \alpha^4(t)} + \frac{3\dot \alpha^3(t)}{8\alpha^5(t)} \bigg ]
    \end{aligned}
\end{equation}
Immediately one can see a difference with the approach from \cite{blaber2020skewed}, as their approximation includes only control velocities rather than acceleration terms. However, for the second-order expression, integration by parts allows us to write the expression entirely in terms of velocities:
\begin{equation}
    \label{eq:next_time_local_boundary}
    \langle W_\text{ex}^{(2),\text{tl}} \rangle = \bigg\{\int_0^{t_f} dt \frac{\dot \alpha^3(t)}{8\alpha^5(t)} \bigg\} - \bigg [ \frac{\dot \alpha^2(t)}{16 \alpha^4(t)} \bigg]_0^{t_f}.
\end{equation}
which recovers the result from \cite{blaber2020skewed} by assuming the protocol begins and ends with 0 velocity, such that the boundary term can be ignored. Integration by parts can similarly be applied to the general case to find
\begin{equation}
    \begin{aligned}
        \frac{1}{4}\zeta^{(2)}_{abc} & = \frac{1}{2} \frac{\partial}{\partial\lambda^c} \bigg(\int_0^{\infty} d\tau \, \tau \, \Psi^{(1)}_{ab}(\tau) \bigg ) \\ & + \int_0^\infty d\tau_1 \int_0^\infty d\tau_2  \Psi_{abc}^{(2)}(\tau_1, \tau_2).
    \end{aligned}
\end{equation}

As discussed by Blaber and Sivak, $\zeta_{ab}^{(1)} + \frac{1}{4}\zeta_{abc}^{(2)} \dot \lambda^c$ defines a Finsler metric on the space of thermodynamic states. However, expressions of this type, which depend only on control parameter velocities cannot be derived beyond this next-order term, as for higher orders acceleration and higher derivatives can no longer all be eliminated in favor of velocity terms and endpoint contributions using integration by parts. The higher-order series mentioned in \cite{Wadia_2022} contains only velocities. Such an expansion can only be obtained if one ignores all acceleration and higher-order derivative terms. However, doing so may lead to inaccurate results as higher-order derivative terms from lower Volterra orders will contribute at the same time-order as the velocity-only term from higher Volterra orders, as shown in App.~\ref{app:drazin_equivalence}. 

Here, we leave the optimization of the higher-order time-local series to future work, since higher-derivative terms require additional endpoint conditions beyond fixed initial and final control values. Additionally, as shown in App.~\ref{app:harmonic}, the linear term in the time-local series diverges for protocols with endpoint jumps, which include a broad class of known exact optimal protocols \cite{seifert2007, zhong2024, aurell2011optimal, Aurell2012BoundaryLayers}, including both the harmonic case and the discrete case considered in App.~\ref{app:discrete_state}. This divergence does not apply for the time-nonlocal approximation Eq.~\eqref{eq:Wex_tot}.

\begin{figure*}[t]
    \centering
    \includegraphics[width=0.87\linewidth]{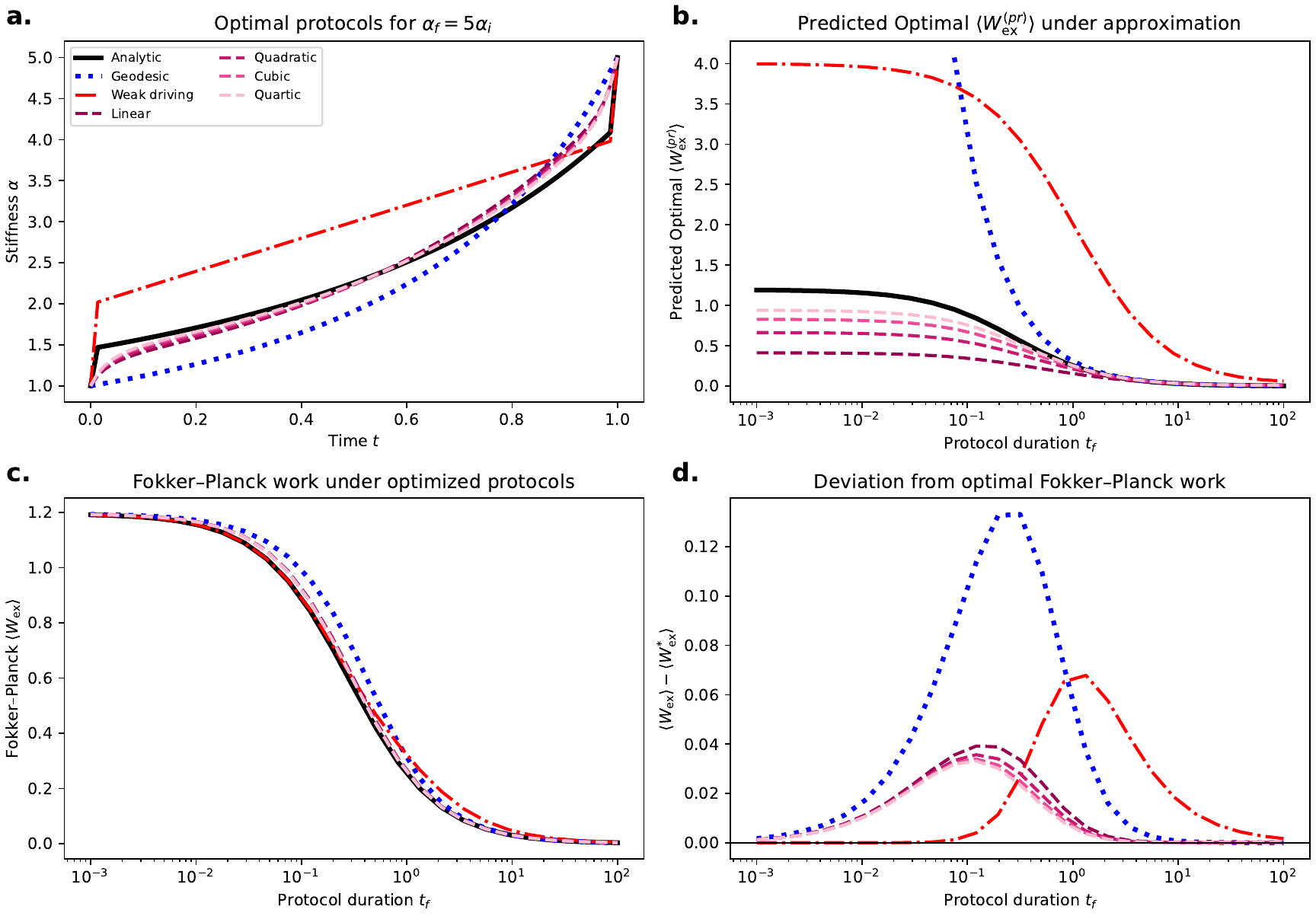}
    \caption{\textit{Comparison of different approximations for the stiffening harmonic oscillator.} Solid black = true optimal protocol from \cite{seifert2007}, blue dotted = Sivak-Crooks geodesic optimal protocol  [Eq.~\eqref{eq:geo_prot}], red dot-dash = Bonança-Deffner weak-driving optimal protocol [Eq.~\eqref{eq:weak_lin_stiff_work}], purple-to-pink dashed curves denote unrestricted linear through quartic response [Eqs.~\eqref{eq:linear_stiff_work} and~\eqref{eq:nth_stiff_work}]. All protocols are computed for a time discretization of 75 steps and a factor of 5 increase in the trap stiffness $\alpha_f = 5 \alpha_i$. a. Optimal protocols for $t_f=1$ according to the different approximations. b. For different protocol durations, the predicted excess work $\langle W_\text{ex}^{(pr)}\rangle$ according to the approximations from Eqs.~\eqref{eq:linear_stiff_work},~\eqref{eq:weak_lin_stiff_work},~\eqref{eq:thermo_geo_stiff_work}, and 
   ~\eqref{eq:nth_stiff_work}, after optimizing the protocol according to the corresponding approximation. c. True excess work $\langle W_\text{ex}\rangle$ for optimal protocols suggested by each approximation at a given protocol duration, evaluated by numerically solving the Fokker-Planck equation. d. Deviation of the true excess work for approximated optimal protocols from true excess work from exact optimal protocol $\langle W_\text{ex}^*\rangle$, as evaluated by numerically solving the Fokker-Planck equation. Note the good performance of weak driving protocols for short times even for strong protocols (despite poor match between the approximated work as shown in panel 4.b.), the good performance of the slow-driving approximations for long times, and the modest improvements of higher-order response terms over linear response in the deviation from the analytically optimal excess work.}
    \label{fig:harmonic-protocols-performance}
\end{figure*}

\subsection{Inclusion of Higher-order Terms}
To evaluate how higher-order terms in the Volterra series affect optimal protocols, we can optimize with respect to a loss function which only includes relaxation functions up to a certain order. This can be done by directly optimizing Eq.~\eqref{eq:Wex_tot}, or by applying the additional weak driving or slow driving approximations. 

In \cite{blaber2020skewed}, Blaber and Sivak additionally point out that unlike the first order contribution, their higher order expression is no longer necessarily nonnegative, allowing for the possibility of unphysical negative excess work and corresponding convergence issues for sufficiently large negative control velocities. This issue is present not only for the next order time-local approximation, but for the higher-order Volterra series for the excess work [Eq.~\eqref{eq:Wex_tot}] more generally. This loss of positivity is a structural property of the truncated higher-order work functional. The $n^{th}$ order contribution in Eq.~\eqref{eq:Wex_tot} contains $(n+1)$ factors of the control velocity. For the harmonic relaxation functions considered here, which are nonnegative given their time ordering, we can say that for protocols which increase stiffness monotonically the excess work remains positive. For decreasing stiffness, however even-order contributions are negative, and can predict arbitrarily negative excess work for sufficiently fast protocols, which can drive optimization instability. Therefore, in our numerical examples we focus on increasing stiffness where optimal protocols are monotonic with non-negative velocity, which avoid the negative-work instability in the optimization (although beyond requiring greater final stiffness, monotonicity is not explicitly enforced - protocols are arrived at by unrestricted optimization).

For the stiffening harmonic oscillator, with $\alpha(t_f) = \alpha_f, \, \alpha(0) = \alpha_i$, we can compare all three approximations, the unconstrained, the weak-driving, and the time-local slow-driving approximations, at first order. These can be written as
\begin{equation}
    \label{eq:linear_stiff_work}
    \langle W^{(1)}_\mathrm{ex} \rangle = \int^{t_f}_{0} \dt \, \dot{\alpha}(t)\int^{t}_{0} \dt_1 \dot{\alpha}(t_1) \Big[ \frac{e^{-2\alpha(t)[t-t_1]}}{2\alpha^2(t)} \Big].
\end{equation}

In the weak driving limit, we have
\begin{equation}
    \label{eq:weak_lin_stiff_work}
    \langle W^{(1)}_\mathrm{ex, \, weak} \rangle = \int^{t_f}_{0} \dt \, \dot{\alpha}(t)\int^{t}_{0} \dt_1 \dot{\alpha}(t_1) \Big[ \frac{e^{-2\alpha_i[t-t_1]}}{2\alpha_i^2} \Big].
\end{equation}
Optimal protocols can be obtained numerically from each of these expressions by discretizing time and varying the control parameter at each point, as described in Appendix~\ref{app:numerical}. From the first order time-local series we have Eq.~\eqref{eq:smooth}.
\begin{equation}
    \label{eq:geo_work}
    \langle W^{(1),\text{tl}}_\mathrm{ex} \rangle = \int^{t_f}_{0} \dt \, \dot{\alpha}^2(t) \Big[ \frac{1}{4\alpha^3(t)}\Big],
\end{equation}
from which the optimal $\alpha(t)$ can be derived analytically
\begin{equation}
    \label{eq:geo_prot}
    \alpha_\mathrm{geo}(t) = \frac{t_f^2 \alpha_i \alpha_f}{(t \sqrt{\alpha_i} + (t_f - t)\sqrt{\alpha_f })^2},
\end{equation}
which gives us
\begin{equation}
    \label{eq:thermo_geo_stiff_work}
    \langle W^{(1),\text{tl}}_\mathrm{ex, \, geo} \rangle = \frac{(\sqrt{\alpha_i} - \sqrt{\alpha_f})^2}{t_f\alpha_f\alpha_i}.
\end{equation}

For the higher-order approximations the approximation for the excess work can be written as
\begin{widetext}
\begin{equation}
    \label{eq:nth_stiff_work}
    \begin{aligned}
        \sum_{n=1}^N \langle W_\text{ex}^{(n)}\rangle = \sum_{n=1}^N \int_0^{t_f} dt \, \dot \alpha(t) \int_0^{t} dt_1 \, \dot \alpha(t_1) \int_0^{t_1} dt_2 ...\int_0^{t_{n-1}} dt_n \, \dot \alpha(t_n) \cdot n! \cdot\Psi^{(n)}_{\alpha...\alpha}(\alpha(t);t-t_1,...,t-t_n), 
    \end{aligned}
\end{equation}
\end{widetext}
where the $n!$ symmetry factor is included due to the change in upper bound on each integral from being set by $t$ to being set by the previous integration time $t_{k-1}$, if integrating over $t_k$. The protocol can again be varied step by step to optimize this function as discussed in App.~\ref{app:numerical}. 

Results from the optimization of these approximations are shown in Fig.~\ref{fig:harmonic-protocols-performance}. For the stiffening harmonic trap, we see that the inclusion of higher-order corrections leads to an improved approximation for the predicted minimal excess work as shown by Fig.\ref{fig:harmonic-protocols-performance}.b, while resulting in a modest improvement in shape of the optimal control protocol prescribed by the changed cost function, as can be seen in Fig.~\ref{fig:harmonic-protocols-performance}.a in comparing the dashed purple (``Linear'') and dashed higher-order (``Quadratic", ``Cubic", ``Quartic") curves. In terms of true performance of these protocols as evaluated by numerically solving the Fokker-Planck equation, we find only a modest improvement in performance by inclusion of higher order terms, as displayed in Fig.~\ref{fig:harmonic-protocols-performance}.d. At the point of highest divergence from the unrestricted linear result the inclusion of higher order response terms in the cost function yields a control protocol with a ~$9-13\%$ improvement beyond linear for quadratic, ~$13-19\%$ for cubic, and ~$16-23 \%$ for quartic. Additionally, the stiffening harmonic case benefits from positivity of all included relaxation functions, which keeps the cost function positive for increasing protocols. In the discrete-state case considered in App.~\ref{app:discrete_state} we give an example with relaxation functions which can be negative. There, for sufficiently slow protocols the inclusion of higher order terms improves performance, but outside the slow-driving regime their inclusion 
can drive protocols away from the true optimum due to their negativity. 

\begin{table}[htbp]
    \centering 
    \begin{tabular}{|c|c|c|c|c|} 
        \hline
        \multicolumn{5}{|c|}{\textbf{Computation times (seconds) - Harmonic}} \\
        \multicolumn{5}{|c|}{single work evaluation, total optimization time} \\
        \hline
        \toprule Steps ($J$) & Linear & Quadratic & Cubic & Quartic \\ 
        % \hline
        30 & $10^{-5}$, $10^{-2}$ & $10^{-4}$, $10^{-1}$ & $10^{-4}$, $10^{-1}$ & $10^{-3}$, $10^{0}$ \\ 
        50 & $10^{-5}$, $10^{-2}$ & $10^{-4}$, $10^{-1}$ & $10^{-3}$, $10^{0}$ & $10^{-2}$, $10^{2}$ \\ 
        75 & $10^{-5}$, $10^{-1}$ & $10^{-3}$, $10^0$ & $10^{-2}$, $10^{1}$ & $10^{-1}$, $10^{3}$ \\
        100& $10^{-4}$, $10^{-1}$ & $10^{-3}$, $10^{1}$ & $10^{-2}$, $10^{2}$ & $10^{0}$, - \\
        150& $10^{-4}$, $10^{0}$ & $10^{-3}$, $10^{2}$ & $10^{-1}$, - & $10^{1}$, - \\
        \hline
    \end{tabular}
    \caption{\textit{Order-of-magnitude computation times for work evaluations and optimization for a stiffening harmonic oscillator.} Computation times shown for discretized protocols of length 30-150, with entries marked ``-" being prohibitively 
    expensive. Optimization time is computed by initializing the control protocol as linear between $\alpha(0) = 1$ and $\alpha(t_f) = 5$, with a duration of $t_f = 1$, and minimizing the corresponding expression given in App.~\ref{app:numerical}. Times are from evaluation on a 2025 MacBook Air laptop with an Apple M4 processor and 24 GB of RAM.}
    \label{tab:comp_times_harmonic} 
\end{table}

A significant computational cost accompanies the inclusion of each higher-order term, since optimizing Eq.~\eqref{eq:nth_stiff_work} requires the evaluation of a higher-dimensional integral. The algorithm for evaluating the unrestricted linear work (as expressed in Appendix~\ref{app:numerical}) for a given protocol has a time complexity $O(J^2)$ while the $n^{th}$ order work has a time complexity $O(J^{n+1})$, with $J$ being the step number in the discretized protocol. Empirical results for computation time are shown in Table~\ref{tab:comp_times_harmonic}, demonstrating that inclusion of higher order response terms is accompanied by a significant computational cost.

\section{Discussion}

Here we have shown how various approximations commonly used in optimal stochastic thermodynamics relate to one another and can be derived from the general Volterra series expansion, and subsequently how higher-order response may be applied to optimal stochastic thermodynamics. Results from near-equilibrium thermodynamic geometry arise from adding a low-acceleration assumption to linear response, while weak-driving results are given by assuming the response of the system is equal to the system's initial response. These approximations are valid for weak and slow driving regimes because higher orders of the Volterra series contribute only for higher orders of driving speed and perturbation strength.

The results for the harmonic oscillator demonstrate that while including higher-order response can improve predictions for excess work and marginally improve optimal protocol performance, the inclusion of higher-order terms incurs a computational cost from the evaluation of higher-dimensional integrals. These results 
are corroborated by the discrete-state example in App.~\ref{app:discrete_state}. Additionally, the inclusion of higher-order terms allows for the possibility of unphysical negative excess work and convergence issues for optimal protocols, as the cost function is no longer guaranteed to be bounded from below. Similarly to the issues resulting from optimizations including higher-order response, as discussed in App.~\ref{app:drazin_equivalence} and App.~\ref{app:harmonic}, time-local slow-driving expansions can be taken to higher orders. However, similar issues arise relating to the possibility of unphysical negative excess work, and convergence to true optimal protocols may be prevented by the divergence of the first term in the time-local series for protocols with jumps.

An interesting future direction would be to 
apply the higher-order time-nonlocal or time-local series to estimate or bound the error of linear approximations. This may include further work in rigorously establishing the convergence of the series' described here for specific cases. Additionally, 
the improvement in accuracy of the approximated excess work near the slow-driving regime could be applied to better understand the energetic costs of controlling nonlinear systems. 

Ultimately, however, the difficulties encountered in optimizing over higher-order approximations suggest a fundamental issue with relying on approximations that expand around $\peq$ in the calculation of optimal protocols in the pursuit of optimal protocols in Region 3 of Fig.~\ref{fig:approx_regimes}. We know that true optimal protocols include discontinuous jumps \cite{seifert2007, zhong2024, aurell2011optimal, Aurell2012BoundaryLayers}. Since the nonequilibrium probability density $\rho_t$ will not instantaneously react to a change in the Hamiltonian, the approximation $\rho_t \approx\rhoeq_{\lambda(t)}$ is fundamentally inapplicable for understanding optimal protocols in Region 3 of Fig.~\ref{fig:approx_regimes} (neither weak nor slow driving). As demonstrated here, attempting to extend near-equilibrium approximations can generate convergence issues and unphysical results while requiring increasingly complex calculations and added computational costs.

Recent results based on optimal transport circumvent this issue by considering the optimal evolution of the underlying density $\rho_t$ \cite{zhong2024}. For overdamped dynamics, we know that in optimal processes the probability density $\rho_t$ follows the Wasserstein geodesic from optimal transport \cite{ito2024geometric, aurell2011optimal, otto2001geometry, villani2021topics}: through integration by parts in both $x$ and $t$ (see \cite{zhong2024} for details), the excess work [Eq.~\eqref{eq:excess_work}] may be rewritten without an explicit $\dot{\lambda}$ dependence:
\begin{gather}
    \langle W_\mathrm{ex} \rangle_{\Lambda} =  \int_0^{t_f} \Vert \dot{\rho}_t \Vert^2_{_\mathrm{OT}} \dt   + D_\mathrm{KL} (\rho_{t_f} | \rhoeq_{\lambda_f}), 
\end{gather}
where $\Vert \dot{\rho}_t \Vert^2_{_\mathrm{OT}}$ is given by the Benamou-Brenier objective function \cite{ito2024geometric, benamou2000computational}, and $D_\mathrm{KL} (\rho_A | \rho_B) := \int \rho_A \ln (\rho_A / \rho_B) \, \dx$ is the KL-divergence. 

Intriguingly, if the solution to the Fokker-Planck equation [Eq.~\eqref{eq:fokker-planck-eq}] can be written as $\rho_t = \rhoeq_{\gamma(t)}$ for a separate curve $\gamma(t) \in \mathcal{M}$, then the excess work can be expressed exactly with the friction tensor [Eq.~\eqref{eq:first-order-correction}]
\begin{equation}
    \langle W_\mathrm{ex} \rangle_{\Lambda}  = \int_0^{t_f}  \dot{\gamma}^a \dot{\gamma}^b \zeta_{a b} \big( \gamma(t) \big) \dt   + D_\mathrm{KL} (\rhoeq_{\gamma(t_f)} | \rhoeq_{\lambda_f}) \label{eq:excess-work-gamma}
\end{equation}
due to a recently discovered equivalence between thermodynamic geometry and optimal transport \cite{zhong2024}.

Rather than relying on expansions around $\peq$, optimal protocols can be derived for any regime (all Regions 0-3 in Fig.~\ref{fig:approx_regimes}) by solving for a path $\gamma(t) \in \mathcal{M}$ that minimizes Eq.~\eqref{eq:excess-work-gamma}, and constructing the optimal protocol $\lambda(t)$ to ensure $\rho_t \approx \rhoeq_{\gamma(t)}$ via counterdiabatic driving (e.g., \cite{iram2021controlling}). In fact, \cite{zhong2024} showed that through this procedure one obtains the \emph{exact} optimal protocol for driven harmonic oscillator example considered here, at the comparatively low computational cost of finding the thermodynamic-geometry geodesic $\gamma(t)$ that minimizes Eq.~\eqref{eq:excess-work-gamma}. This suggests that expansions around a separate dynamically evolving reference density, 
such as $\rhoeq_{\gamma(t)}$ (c.f., \cite{zhong2024, zhong2025optimal}), rather than the instantaneous equilibrium state 
$\rhoeq_{\lambda(t)}$, may be a more promising route for developing efficient optimal-control methods far from equilibrium.

\section{Acknowledgements}

We would like to thank Neha Wadia, Hana Mir, and Steven Blaber for insightful conversations.
S.D. was supported in part by an H2H8 research fellowship. A.Z. was supported by the Department of Defense (DoD) through the National Defense Science \& Engineering Graduate (NDSEG) Fellowship Program. This research was supported by CoCoSys, one of the seven centers in JUMP 2.0, a Semiconductor Research Corporation (SRC) program sponsored by DARPA; in part by grants from the NSF (DMS-2235451) and Simons Foundation (MPS-NITMB-00005320) to the NSF-Simons National Institute for Theory and Mathematics in Biology (NITMB); as well as in part by the U.S. Army Research Laboratory and the U.S. Army Research Office under Contract No. W911NF-20-1-0151.

\appendix

\section{Relation Between Taylor-expanded Volterra Series and Drazin Inverse Series}

\label{app:drazin_equivalence}

Here we show that Taylor-expanding the time-nonlocal Volterra series and grouping its terms by total temporal-derivative order is equivalent to the recursive Drazin inverse series. We can begin by noting the equivalence between the spectral perturbative series introduced in \cite{Wadia_2022} and recursive application of the Drazin inverse as discussed in \cite{sawchuk2024dynamical}. With the Green's function of the Fokker-Planck operator introduced in \cite{Wadia_2022} satisfying
\begin{equation}
    \mathcal{L}_{\lambda(t)}(x) G_{\lambda(t)}(x,y) = \delta(x-y) - \rhoeq_{\lambda(t)}(x),
\end{equation}
we can see that for any zero-normalization function $g$ (i.e., satisfying $\int dx  \, g(x) = 0$),
\begin{equation}
    \mathcal{L}_{\lambda(t)}^{-1}[g](x) = \int dy \, G_{\lambda(t)}(x,y) g(y). 
\end{equation}
Thus, since the deviation from equilibrium $\delta \rho(x) = \rho_t(x) - \rhoeq_{\lambda(t)}(x)$ has a normalization of zero $\int dx \, \delta \rho (x) = 0$, we can see that the Green's function recursion relation
\begin{equation}
    \delta \rho_t^{(m),\text{tl}}(x) = \int dy \, G_{\lambda(t)}(x,y) \, \partial_t\delta \rho_t^{(m-1),\text{tl}}(y),
\end{equation}
is equivalent to
\begin{equation}
    \delta \rho_t^{(m),\text{tl}}(x) = \mathcal{L}_{\lambda(t)}^{-1}\big [ \partial_t \delta \rho^{(m-1),\text{tl}}_t \big],
\end{equation}
with $\delta \rho_t^{(0),\text{tl}} = \rhoeq_{\lambda(t)}$.

To connect this recursion to the time-nonlocal response series, following \cite{Wadia_2022} we can introduce a slow time
\begin{equation}
    \label{eq:app_rescaling}
    \tilde{t} = \nu t, \,\,\, \lambda(t) = \lambda(\tilde t)
\end{equation}
where $\nu$ allows us to count the order of the approximation around $t$. We can write the $n^{th}$ order correction from the time-nonlocal Volterra series as Eq.~\eqref{eq:nth_order_correction} from the main text:
\begin{equation}
    \label{eq:app_nth_correction}
    \delta \rho_t^{(n)}(x) = \int_{\mathbb{R}^n_+} d^n \tau \, h^{(n)}_{i_1...i_n}(x;\lambda(t), \tau_1,..., \tau_n) \prod^n_{r=1} \dot \lambda^{i_r}(t-\tau_r).
\end{equation}
Here, the factor of $1/n!$ is included in the response function due to the integration over all $\mathbb{R}^n_+$. However, if one restricts the integration region to the ordered region $0 < \tau_1 < ... < \tau_n$, the $1/n!$ must be removed.

Each velocity in each order of the Volterra series has its own Taylor expansion,
\begin{equation}
    \dot \lambda^{i_k}(t-\tau_k) = \sum^{\infty}_{m_k = 0} \frac{(-\tau_k)^{m_k}}{m_k!} \frac{d^{m_k}}{dt^{m_k}} \dot \lambda^{i_k}(t).
\end{equation}
Substituting this into the $n^{th}$ order Volterra term Eq.~\eqref{eq:app_nth_correction} gives
\begin{align}
\delta\rho_t^{(n)}(x)
={}&
\sum^\infty_{m_1,\ldots,m_n\geq0}
\frac{(-1)^{m_1+\cdots+m_n}}
{m_1!\cdots m_n!}
\nonumber\\
&\times
\left[
\int_{\mathbb R_+^n}d^n\tau\,
\prod_{k=1}^{n}\tau_k^{m_k}
h_{i_1\ldots i_n}^{(n)}
\bigl(x;\lambda(t),\tau_1,\ldots,\tau_n\bigr)
\right]
\nonumber\\
&\times
\prod_{k=1}^{n}
\frac{d^{m_k}}{dt^{m_k}}
\dot{\lambda}^{i_k}(t).
\label{eq:app-expanded-volterra}
\end{align}

Introducing the rescaling Eq.~\eqref{eq:app_rescaling}, each delayed velocity becomes
\begin{equation}
    \begin{aligned}
        \dot \lambda^{i_k}(t-\tau_k) & = \nu \frac{d \lambda^{i_k}}{d \tilde t} (\tilde t - \nu \tau_k)\\
        & =\sum_{m_k = 0}^\infty \nu^{m_k+1} \frac{(-\tau_k)^{m_k}}{m_k!} \frac{d^{m_k+1}}{d \tilde t^{m_k+1}} \lambda^{i_k}(\tilde t)
    \end{aligned}
\end{equation}
The parameter $\nu$ allows us to track the $m^{th}$ time-local expansion, with Volterra orders marked by $n$, and Taylor-expansion orders marked by $m_k$. These are given by terms of order
\begin{equation}
    \nu^{n+m_1+...+m_n},
\end{equation}
where
\begin{equation}
    n+m_1+...+m_n = m, \,\,\, 1 \leq n \leq m.
\end{equation}

Substituting the rescaling into Eq.~\eqref{eq:app_nth_correction} we can identify the terms which are proportional to $\nu^m$:
\begin{align}
\label{eq:app_mth_order_time_local}
\delta\rho_{\tilde t}^{(m),\text{tl}}(x)
={}&
\sum_{n=1}^{m}
\sum_{\substack{m_1,\ldots,m_n\geq0\\
n+m_1+\cdots+m_n=m}}
\frac{(-1)^{m-n}}
{m_1!\cdots m_n!}
\nonumber\\
\times & 
\left[
\int_{\mathbb R_+^n}d^n\tau\,
\prod_{k=1}^{n}\tau_k^{m_k}
h_{i_1\ldots i_n}^{(n)}
\bigl(x;\lambda(\tilde t),\tau_1,\ldots,\tau_n\bigr)
\right]
\nonumber\\
\times & 
\prod_{k=1}^{n}
\frac{d^{m_k+1}}{d \tilde t^{m_k+1}}
{\lambda}^{i_k}(\tilde t).
\end{align}
This clarifies that the $m^{th}$ order time-local expansion term is not just from the $m^{th}$ order Volterra kernel. It gets higher order contributions from every Volterra kernel of order $1 \leq n \leq m$. For example, the second time-local order contains both the acceleration term obtained from the first-order Volterra kernel, and the two-velocity term obtained from the second-order Volterra kernel (as shown by Eq.~\eqref{eq:second_order_time_local} in Section.~\ref{sec:higher_time_local_order}). Similarly, the third time-local order contains the third time derivative from the first-order kernel, the velocity-acceleration terms from the second-order kernel, and the three-velocity term from the third-order kernel.

These time local terms can be arranged as an approximation of order $M$:
\begin{equation}
    \label{eq:app_time_local_expansion}
    \rho_{\tilde t} = \rhoeq_{\lambda(\tilde t)} + \sum_{m=1}^M \nu^m \delta \rho_{\tilde t}^{(m),\text{tl}} + O(\nu^{M+1}),
\end{equation}
which can be inserted into the rescaled time evolution equation
\begin{equation}
    \nu \partial_{\tilde t} \rho_{\tilde t}= \mathcal{L}_{\lambda(\tilde t)} \rho_{\tilde t}.
\end{equation}
From this we obtain
\begin{equation}
    \begin{aligned}
        \nu \partial_{\tilde t} \rhoeq_{\lambda(\tilde t)} &+ \sum_{m=1}^{M} \nu^{m+1} \partial_{\tilde t} \delta \rho_{\tilde t}^{(m),\text{tl}} \\ = &\sum_{m=1}^{M} \nu^m \mathcal{L}_{\lambda(\tilde t)} \delta \rho_{\tilde t}^{(m),\text{tl}} + O(\nu^{M+1}),
    \end{aligned}
\end{equation}
where we have used $\mathcal{L}_{\lambda(\tilde t)} \rhoeq_{\lambda(\tilde t)} = 0$. Matching equal powers of $\nu$ we find
\begin{equation}
    \mathcal{L}_{\lambda(\tilde t)} \delta \rho_{\tilde t}^{(1),\text{tl}} = \partial_{\tilde t} \rhoeq_{\lambda(\tilde t)},
\end{equation}
and
\begin{equation}
    \mathcal{L}_{\lambda(\tilde t)} \delta \rho_{\tilde t}^{(m),\text{tl}} = \partial_{\tilde t} \delta \rho^{(m-1),\text{tl}}_{\tilde t},
\end{equation}
with $\int dx\, \delta \rho^{(m),\text{tl}}_{\tilde t}(x) = 0$ for all $m \geq 1$.
This allows us to identify the Drazin inverse or Green's function series with a Taylor expansion of the velocities found in the Volterra series:
\begin{equation}
    \delta \rho^{(m),\text{tl}}_{\tilde t} = \mathcal{L}^{-1}_{\lambda(\tilde t)} \partial_{\tilde t} \delta \rho^{(m-1),\text{tl}}_{\tilde t},
\end{equation}
where $\delta \rho^{(m),\text{tl}}_{\tilde t}$ is given by Eq.\eqref{eq:app_mth_order_time_local}.

Additionally, we can identify the explicit relationship between the quantities in the square brackets in Eq.~\eqref{eq:app_mth_order_time_local}, which are temporal moments of the response kernels, and the Drazin inverse. For the ordered kernel $\tau_1 < ... < \tau_n$, the expression depends on successive time gaps, $\tau_1$, $\tau_2 - \tau_1$, ..., $\tau_n - \tau_{n-1}$. Each gap is nonnegative and appears in the time-evolution operators generated in each Volterra kernel. Referring to the successive gaps as $s_1=\tau_1$ and $s_j=\tau_j-\tau_{j-1}$, we can see $\tau_k=s_1+\cdots+s_k$. Expanding the factors $\prod_k\tau_k^{m_k}$ produces sums of products $\prod_js_j^{q_j}$. The identity
\begin{equation}
    \label{eq:higher_drazin}
    \int_0^\infty ds_j \, s_j^{q_j} e^{\mathcal{L}_{\lambda(t)}s_j} g = (-1)^{q_j+1}q_j! \big [ \mathcal{L}^{-1}_{\lambda(t)}\big]^{q_j+1} g.
\end{equation}
can then be applied successively to each gap integral, giving us ordered products of powers of \(\mathcal L_\lambda^{-1}\) separated by the parameter derivatives appearing in the Volterra kernel. This identity can be justified by observing that since $g$ has a normalization of zero, we decompose it in terms of non-stationary eigenmodes of $\mathcal{L}_{\lambda(t)}$, such that $g = \sum c_r \, g_r$, where $\mathcal{L}_{\lambda(t)} g_r = r g_r$ and $r < 0$, then
\begin{equation}
    \begin{aligned}
    \int_0^\infty & \,ds_j \, s_j^{q_j} e^{rs_j} g_r = \frac{q_j!}{(-r)^{q_j+1}}g_r \\ = &(-1)^{q_j+1}q_j! \bigg ( \frac{1}{r} \bigg)^{q_j+1}g_r 
    \end{aligned}
\end{equation}
The completeness of the eigenmodes gives us Eq.~\eqref{eq:higher_drazin} for any zero-normalization $g$.

\section{Higher-Order Time-local Expansion for Moving Overdamped Harmonic Trap}
\label{app:harmonic}

Following the approach indentified in \cite{Wadia_2022}, we can assess the potential of the higher-order time-local approach by focusing on the simple case of a moving harmonic trap of fixed stiffness. The time-local expansion for the probability density can be written as
\begin{equation}
    \label{eq:perurbative_expansion}
    \begin{aligned}
        & \rho(x;t) = \rhoeq_{\lambda(t)}(x) + \int  \dx' G_{\lambda(t)}(x;x') \frac{\partial}{\partial t} \rho_{\lambda(t)}^\eq(x') \\
        & + \int  \dx'' G_{\lambda}(x;x'') \frac{\partial}{\partial t} \int  \dx' G_{\lambda(t)}(x'';x') \frac{\partial}{\partial t} \rhoeq_{\lambda(t)}(x') + ...
    \end{aligned}
\end{equation}
This expression can be simplified by writing it in terms of an integral operator
\begin{equation}
    \hat{G}_{\lambda(t)} = \int dy \, G_{\lambda(t)}(x;y) = \int dy \, \sum_{n \geq 1} \frac{\rho_{r,n}(x)\rho_{l,n}(y)}{r_n}.
\end{equation}
As with Eq.~\eqref{eq:recursive_drazin}, written in terms of the integral operator the $k^{th}$ term in the time-local series can be written as follows, where $\rho^{(0)}_{\lambda(t)}(x) = \rhoeq_{\lambda(t)}(x)$:
\begin{equation}
    \rho^{(k)}_{\lambda(t)}(x) = \Big ( \hat{G}_{\lambda(t)} \frac{d}{dt} \Big )^k \rhoeq_{\lambda(t)}(x).
\end{equation}
By inserting Eq.~\eqref{eq:perurbative_expansion} into the original definition of work, Eq.~\eqref{eq:total_work}, where the power $\frac{d\langle W \rangle}{dt} \approx \frac{d\langle W_{rev} \rangle}{dt} + \sum_{k = 1} \frac{d\langle W^{(k)P}_\text{ex} \rangle}{dt}$, we arrive at
\begin{equation}
    \frac{d\langle W^{(k),\text{tl}}_\text{ex} \rangle}{dt} = \int  \dx \frac{dH_{\lambda(t)}(x)}{dt} \Big ( \hat{G}_{\lambda(t)} \frac{d}{dt} \Big )^{k} \rhoeq_{\lambda(t)}.
\end{equation}
Here we assume the same potential as Eq.~\eqref{eq:SHO_potential}, giving us the same equilibrium distribution and left eigenfunction Eq.~\eqref{eq:left-eigenvector}, but with $\alpha = 1$: 
\begin{equation}
    V_{\mu(t)}(x) = \frac{(x - \mu(t))^2}{2}.
\end{equation}
We can then consider the successive application of the time-derivative and Green's function integral operators to the equilibrium distribution to evaluate higher-order terms in the work expansion. To do so, let us first point out some important relationships. Recalling $\rhoeq_{\lambda(t)} = e^{-H_{\lambda(t)}(x)}/Z$, for the moving harmonic oscillator with static stiffness, the partition function $Z$ is time independent
\begin{equation}
    Z(t) = \int  \dx \, e^{-[x-\mu(t)]^2/2} = \sqrt{2\pi}.
\end{equation}

This observation simplifies further steps, which might otherwise be complex. The eigenfunctions are given by
\begin{equation}
    \rho_{r,n} = \rhoeq_{\mu(t)} \rho_{l,n} = \frac{1}{\sqrt{n!}} \text{He}_n(x-\mu(t)) \rhoeq_{\mu(t)},
\end{equation}
which correspond to eigenvalues
\begin{equation}
    r_n = -n.
\end{equation}

Noticing $\dot{V}_{\lambda(t)} = d V_{\lambda(t)} / \dt = -\dot{\mu}[x-\mu] = \dot{\mu} \rho_{l,1}$, and $d\rhoeq_{\mu(t)}/dt = -\dot{V}_{\mu(t)} \rhoeq_{\mu(t)} = \dot{\mu}[x-\mu] \rhoeq_{\mu(t)} = \dot{\mu} \rho_{r,1}$, we can evaluate the first term of the series 
\begin{equation}
    \label{eq:perturbative-first-order}
    \begin{aligned}
        & \frac{d \langle W^{(1),\text{tl}}_\text{ex} \rangle}{dt} = \int  \dx \, \dot{V}_{\mu(t)} \Big ( \hat{G}_{\mu(t)} \frac{d}{dt} \Big ) \rhoeq_{\mu(t)} \\
        & = -\int  \dx \, \dot{\mu}(t) \, \rho_{l,1}(x) \int dy \, \sum_{n \geq 1} \frac{\rho_{r,n}(x) \rho_{l,n}(y)}{r_n} \, \dot{\mu}(t) \, \rho_{r,1}(y). \\
    \end{aligned}
\end{equation}
Applying Eq.~\eqref{eq:orthonormal}, we see that
\begin{equation}
    \frac{d \langle W^{(1),\text{tl}}_\text{ex} \rangle}{dt} = -\frac{\dot{\mu}^2(t)}{r_1}.
\end{equation}

Higher-order terms can be evaluated by noticing that, by Eq.~\eqref{eq:orthonormal}, only terms with $\rho_{r,1}$ will remain. We can derive a simple expression for this expansion to all orders by using the fact that the terms with $\rho_{r,n}$ for $n \geq 2$ vanish, by considering the application of the operator $\hat{G}_{\mu(t)} \frac{d}{dt}$ to the product of a function of $\mu$ and its derivatives, $\phi(\mu, \dot{\mu},\ddot{\mu}, ...) = \phi(t)$ with right eigenfunction $\rho_{r,k}$ 
\begin{equation}
    \label{eq:Ghatdtphi}
    \begin{aligned}
        &\hat{G}_{\lambda(t)} \frac{d}{dt}\Big( \phi(t) \rho_{r,k} \Big) = \hat{G}_{\lambda(t)} \frac{d}{dt}\Big( \dot{\phi}(t) \rho_{r,k} \\
        & + \phi(t) \frac{-\dot{\mu}(t) \text{He}_k'(x-\mu(t))}{\sqrt{k!}} \rhoeq_{\mu(t)} \\
        & + \phi(t) \frac{\dot{\mu}(t)[x-\mu(t)] \text{He}_k(x-\mu(t))}{\sqrt{k!}} \rhoeq_{\mu(t)} \Big). 
    \end{aligned}
\end{equation}

Using the Hermite polynomial identity
\begin{equation}
    \text{He}_{n+1}(x)=x\text{He}_n(x) - \text{He}_n'(x),
\end{equation}
Eq.~\eqref{eq:Ghatdtphi} becomes
\begin{equation}
    \hat{G}_{\lambda(t)} \frac{d}{dt}\Big( \phi(t) \rho_{r,k} \Big) = \frac{\dot{\phi}(t)}{r_k} \rho_{r,k} + \frac{\dot{\mu}(t)\phi(t)\sqrt{k+1}}{r_{k+1}} \rho_{r,k+1}.
\end{equation}
The first order term in the series is proportional to $\rho_{r,1}$ as seen in the second line of Eq.~\eqref{eq:perturbative-first-order}. For second- and higher-order terms, only those that are proportional to $\rho_{r,1}$ remain, while all terms proportional to $\rho_{r,n}$ with $n > 1$, vanish. This gives us the general expression for all orders:
\begin{equation}
    \label{eq:all_orders_work}
    \begin{aligned}
        \frac{d\langle W^{(k),\text{tl}}_\text{ex} \rangle}{dt} & = -\dot{\mu}(t)\Big(\frac{d}{dt}\Big)^k \frac{[\mu(t)]}{\prod^k_{i=1} r_1}  \\
        = -\dot{\mu}(t) & \Big ( \frac{d}{dt} \Big )^k \frac{\mu(t)}{(-1)^k} = -\dot{\mu}(t) \Big (-\frac{d}{dt} \Big )^k \mu(t).
    \end{aligned}
\end{equation}
giving us the work to an arbitrary expansion order $K$,
\begin{equation}
    \label{eq:work_pert_all_orders}
    \langle W_\text{ex} \rangle \approx \int_0^{t_f} \dt \, \Big[ \sum^K_{k=1} -\dot\mu(t) \Big ( - \frac{d}{dt}\Big )^{k} \mu(t) \Big].
\end{equation}

The full expansion allows us to identify two issues arising from the inclusion of higher orders of the time-local expansion. First, the even-$k$ higher orders will contribute negative work, meaning that if the series is truncated at the $K$th order, the function $\mu(t)$ can be optimized to maximize odd orders and minimize even orders, allowing for the possibility of unphysical negative excess work. The work truncated at quadratic order $K = 2$ can be written as

\begin{equation}
    \label{eq:pert_next_od_work}
    \begin{aligned}
    \langle W_\mathrm{ex} \rangle & \approx \int_0^{t_f} \dt \, \Big(\frac{d\langle W^{(1),\text{tl}} \rangle}{dt} + \frac{d\langle W^{(2),\text{tl}} \rangle}{dt} \Big )\\
    & = \int^{t_f}_0 \dt \, \Big ( \dot{\mu}^2(t) - \dot{\mu}(t) \ddot{\mu}(t) \Big ).
    \end{aligned}
\end{equation}
For even a simple protocol we can obtain a negative excess work:
\begin{equation}
    \mu(t) =
        \begin{cases}
        0, & t < 0, \\
        \frac{t^2}{2}, & 0 \le t \le t_f, \\
        \frac{t_f^2}{2}, & t > t_f .
        \end{cases}
\end{equation}
Plugging this into Eq.~\eqref{eq:pert_next_od_work} and integrating we find
\begin{equation}
    \langle W_\mathrm{ex} \rangle \approx \langle W_\text{ex}^{(1),\text{tl}} \rangle + \langle W^{(2),\text{tl}}_\text{ex} \rangle =  t^2_f \Big ( \frac{t_f}{3} - \frac{1}{2} \Big ),
\end{equation}
which is only non-negative for $t_f > 3/2$. Unphysical negative excess work is recovered in the low $t_f$ weak and fast limit. Here we assume that the endpoint derivatives are given by their interior values. Eq.~\eqref{eq:next_time_local_boundary} instead makes the assumption that the endpoint derivatives are zero, however, the resulting approximation can also lead to unphysical negative work as discussed in \cite{blaber2020skewed}.

Second, we can see that relying on this series can prevent the recovery of optimal protocols since the first term in the series diverges for protocols with jumps, which are present for the moving harmonic trap and common in optimal protocols generally \cite{seifert2007, zhong2024, aurell2011optimal, Aurell2012BoundaryLayers}. The true optimal protocol for a moving harmonic trap starting at $\mu_i = 0$ is
\begin{equation}
    \mu^*(t) =
        \begin{cases}
        0, 
        & t < 0, \\[10pt]
        \dfrac{\mu_f(t+1)}{t_f+2},
        & 0 \leq t \leq t_f, \\[6pt]
        \mu_f,
        & t \ge t_f,
        \end{cases}
\end{equation}
which exhibits jumps of size $\mu_f/(t_f+2)$ at its endpoints \cite{seifert2007}. This gives us the optimal velocity
\begin{equation}
    \label{eq:moving_velocity}
    \dot \mu^*(t) =  \frac{\mu_f}{t_f+2} \cdot\bigg [\delta(t) + (\Theta(t) - \Theta(t-t_f))  + \delta (t-t_f) \bigg ].
\end{equation}

Integrating over $(\dot \mu^*)^2$ is simplified by evaluating the $\delta$ and $\Theta$ functions, which appear with different time arguments. The first order of excess work is (with $0^-$ and $t_f^+$ specifying that the integral is taken over the entirety of the delta functions)
\begin{equation}
    \begin{aligned}
        \label{eq:delta_work}
        \langle W^{(1),\text{tl}}_\text{ex} \rangle &= \int^{t_f^+}_{0^-} dt \, \frac{\mu_f^2}{(t_f+2)^2} \bigg [ \delta(t) \\ &+ \Theta(t) - \Theta(t-t_f) + \delta(t-t_f) \bigg ]^2.
    \end{aligned}
\end{equation}

We can investigate this expression by regularizing the Dirac delta function as a normal distribution of width $\varepsilon$, 
\begin{equation}
    \label{eq:reg_delta}
    \delta(t) = \lim_{\epsilon \rightarrow 0} \frac{1}{\sqrt{2\pi} \varepsilon} \exp \bigg [- \frac{t^2}{2\varepsilon^2} \bigg].
\end{equation} 
Inserting this into Eq.~\eqref{eq:delta_work}, we find that for the optimal protocol, and taking the limit as $\epsilon \rightarrow 0$ the first term in the series is divergent: 
\begin{equation}    
    \label{eq:divergent}
    \begin{aligned}
        \langle W^{(1),\text{tl}}_\text{ex} \rangle & = \frac{\mu_f^2}{(t_f+2)^2} \bigg [ t_f + 2 +\lim_{\varepsilon \rightarrow 0}\int^{\infty}_{-\infty} dt \, \bigg [ \frac{e^{-t^2/\varepsilon^2}}{2\pi \varepsilon^2} \\ & \ \ \  + \frac{e^{-(t - t_f)^2/\varepsilon^2}}{2\pi \varepsilon^2} \bigg ] \\
        & = \frac{\mu_f^2}{(t_f+2)^2} \bigg [ t_f + 2+ \lim_{\varepsilon \rightarrow 0}\frac{1}{\sqrt{\pi\varepsilon}}\bigg ] = \infty.
    \end{aligned}
\end{equation}
Because of this, we should not expect the time-local series to converge to true optimal protocols, which are often functions with discontinuities.

This divergence is not present for the time-nonlocal approximation. Adopting Eq.~\eqref{eq:Wex1} as our objective function, we can similarly evaluate linear term in the work expansion Eq.~\eqref{eq:Wex_tot} with respect to the true optimal protocol Eq.~\eqref{eq:moving_velocity}. First we can evaluate the relaxation function in this case, which is
\begin{equation}
    \Psi_{\mu\mu}^{(1)}(\mu(t);t-t_1) = e^{-(t-t_1)}.
\end{equation}
Inserting this into Eq.~\eqref{eq:Wex1} we see
\begin{equation}
    \label{eq:Wex_1_moving}
    \langle W_\text{ex}^{(1)} \rangle = \int_{0^-}^{t_f^+}dt \, \dot \mu^* (t) \int_{-\infty}^{t} dt_1 \, e^{-(t-t_1)} \dot \mu^*(t_1).
\end{equation}
With $\mu^{*}(t)$ given by Eq.~\eqref{eq:moving_velocity}, with the inclusion of the regularized delta function Eq.~\eqref{eq:reg_delta}, Eq.~\eqref{eq:Wex_1_moving} becomes
\begin{equation}
    \begin{aligned}
        \langle W_\text{ex}^{(1)} \rangle & = \frac{\mu_{f}^2}{(t_f+2)^2} \lim_{\varepsilon \rightarrow0} \int_{-\infty}^{\infty} dt \bigg[ \frac{e^{-t^2/2\varepsilon^2}}{\sqrt{2\pi} \varepsilon} + \Theta(t) - \Theta(t-t_f) \\ & + \frac{e^{-(t-t_f)^2/2\varepsilon^2}}{\sqrt{2\pi} } \bigg ] \int_{-\infty}^{t_1}dt_1 \, e^{-(t-t_1)} \bigg[ \frac{e^{-t_1^2/2\varepsilon^2}}{\sqrt{2\pi} \varepsilon} \\ & + \Theta(t_1) - \Theta(t_1-t_f) + \frac{e^{-(t_1-t_f)^2/2\varepsilon^2}}{\sqrt{2\pi} } \bigg ] \\ & = \frac{\mu_f^2}{(t_f+2)^2} \bigg [ (t_f+2) - \lim_{\varepsilon \rightarrow 0}\frac{2\sqrt{\varepsilon}}{\sqrt{\pi}} + O(\varepsilon^2) \bigg] \\ & = \frac{\mu_f^2}{(t_f+2)},
    \end{aligned}
\end{equation}
which we can see converges as $\varepsilon \rightarrow 0$. In fact, this expression is the exact optimal work found in \cite{seifert2007}. This is because the higher order response functions for the moving (no stiffening) harmonic trap vanish, and in the bulk of the protocol the acceleration and higher order derivatives beyond velocity are zero, meaning that Eq.~\eqref{eq:Wex_1_moving} captures all contributions to the work.

\section{Numerical Optimization and Work Evaluation for Stiffening Harmonic Trap}

\label{app:numerical}

To numerically optimize the expressions in Section \ref{sec:harmonic}, protocols were discretized in time to $J=75$ equal time intervals between $t = 0$ and $t = t_f$ and optimized by varying the position of each of these 100 points. Protocols are initialized as linear between initial and final control parameters. Integrals were evaluated as discrete sums using the left hand rule, corresponding to an It\^{o}
 interpretation of the stochastic dynamics. 
 
Following Eq.~\eqref{eq:nth_stiff_work} we integrate over the ordered region $t > t_1 >... > t_n$ and include the symmetry factor $n!$. This provides a computational speedup by restricting the multi-dimensional integration domain. In order to properly discretize the multi-dimensional integrals, care must be taken to compute the accessible volume within individual bins that straddle the boundary defined by limits of integration. For example, with discrete time steps $j_n$, terms for which either $j_0 = j_1$ or $j_1 = j_2$, without all three being equal, are multiplied by a factor of $1/2$, terms for which $j_0 = j_1 = j_2$ are multiplied by a factor of $1/6$. In general, if $B$ is the set of maximal blocks of equal indices, and $q_b$ is the number of equal indices in each block, this can be expressed by the factor
 \begin{equation}
     w(j_0,...,j_n) = \prod_{b\in B} \frac{1}{q_b!}.
 \end{equation}
 
 For example, for a 3-dimensional integral we would have
 
 \begin{equation}
     w(j_0,j_1,j_2) = \begin{cases}
         1, \ \ \ \ \ \ j_0 > j_1 > j_2 \\
         1/2, \ \ \ j_0 = j_1 > j_2 \text{ or } j_0 > j_1 = j_2 \\
         1/6, \ \ \ j_0 = j_1 = j_2
     \end{cases}.
 \end{equation}
 
To approximate the excess work to linear order Eq.~\eqref{eq:linear_stiff_work} becomes (with $J = 75$, $\Delta t = t_f/J$):

\begin{equation}
    \label{eq:disc-lin}
    \begin{aligned}
        \langle W^{(1)}_\mathrm{ex} \rangle & \approx \sum^{J-1}_{j_0=0} \Delta t \frac{\alpha_{j_0+1} - \alpha_{j_0}}{\Delta t} \\ \cdot & \sum^{j_0}_{j_1=0} \Delta t \frac{\alpha_{j_1+1} - \alpha_{j_1}}{\Delta t} \frac{w(j_0,j_1)e^{-2\alpha_{j_0} \Delta t (j_0-j_1)}}{2 \alpha^2_{j_0}}.
    \end{aligned}
\end{equation}
For the weak driving limit, we set $\alpha_{j_0}$ in the response function to $\alpha_0$, 
so that the left hand side of Eq.~\eqref{eq:weak_lin_stiff_work} could be approximated as 

\begin{equation}
    \label{eq:disc-weak}
    \begin{aligned}
        \langle W^{(1)}_\mathrm{ex} \rangle & \approx \sum^{J-1}_{j_0=0} \Delta t \frac{\alpha_{j_0+1} - \alpha_{j_0}}{\Delta t} \\ \cdot & \sum^{j_0}_{j_1=0} \Delta t \frac{\alpha_{j_1+1} - \alpha_{j_1}}{\Delta t} \frac{w(j_0,j_1)e^{-2\alpha_{0} \Delta t (j_0-j_1)}}{2 \alpha^2_{0}}.
    \end{aligned}
\end{equation}

Approximating the excess work to $N^{th}$ order, Eq.~\eqref{eq:nth_stiff_work} becomes
\begin{widetext}
\begin{equation}
    \label{eq:disc-quad}
    \begin{aligned}
        \sum_{n=1}^N \langle W_\text{ex}^{(n)}\rangle & \approx \sum_{n=1}^N\sum^{J-1}_{j_0=0} \sum_{j_1=0}^{j_0} ... \sum_{j_n = 0}^{j_{n-1}} \prod_{r=0}^n \bigg [ \Delta t\frac{\alpha_{j_r+1} - \alpha_{j_r}}{\Delta t} \bigg] \cdot w(j_0,...,j_n) \cdot n! \cdot \Psi_{\alpha...\alpha}^{(n)}(\alpha_{j_0}; \Delta t (j_0-j_1), ... , \Delta t (j_{0} -j_n)).
    \end{aligned}
\end{equation}
\end{widetext}

Each of these expressions is minimized by varying $[\alpha_1, \alpha_2, ... , \alpha_{J-1}]$, starting from a linear protocol. The optimal protocol for a given approximation is determined by minimizing Eqns.~\eqref{eq:disc-lin}, \eqref{eq:disc-weak}, and \eqref{eq:disc-quad} using \texttt{scipy.optimize.minimize} \cite{2020SciPy-NMeth} according to the corresponding assumptions.

For each of these approximations, we compared the values given by optimizing them with the true values for the work, which we found by solving the Fokker-Planck equation numerically via the discretization in \cite{holubec2019physically, zhong2022limited}. For this, protocols optimized according to the above approximations were fed into a Fokker-Planck simulation with a lattice-discretized state space with spacing $\Delta x = 0.1$ from $x = -5$ to $x = 5$. Code found at: 
\url{https://github.com/SamuelDAmbrosia/Higher-order-stochastic-thermo}. 

\section{Optimal Control of a Discrete-State Double Well}

\label{app:discrete_state}

We can demonstrate these results for a discrete system by focusing on a discrete approximation of a double well potential, with two stable states separated by an energy barrier. We can label these stable states 0 and 1, such that $x \in \{0,1\}$. The two states have energies $E_1$ and $E_0$, and the time dependent probability distribution over them is given by
\begin{equation}
    p(t) = \begin{pmatrix}
        p_0 (t) \\
        p_1 (t)
    \end{pmatrix}, \,\, p_0 + p_1 = 1.
\end{equation}

The dynamics of the system are specified by a rate matrix $K_{\lambda(t)}$, fixed by control parameters $\lambda(t)$, by
\begin{equation}
    \label{eq:master}
    \dot p (t) = K_{\lambda(t)} p (t).
\end{equation}
$K_{\lambda(t)}$ can be fixed by using an Arrhenius equation for the double well potential. If there is a barrier of energy $\Phi$ between 0 and 1, and we fix $E_0 = 0$ and $E_1 = \upsilon$, the transition rates between the states are $K_{0\to 1} = A e^{-\beta \Phi}$ and $K_{1\to 0} = A e^{-\beta (\Phi - \upsilon)}$, where $A$ is the Arrhenius factor. Combining constants as $\Gamma = A e^{-\beta \Phi}$, the rate matrix is
\begin{equation}
    K_{\lambda(t)} = \begin{pmatrix}
        -\Gamma & \Gamma e^{\beta \upsilon(t)} \\
        \Gamma & - \Gamma e^{\beta \upsilon(t)}
    \end{pmatrix},
\end{equation}
which satisfies the detailed-balance condition
\begin{equation}
    \frac{K_{0\to 1}}{K_{1\to 0}} = e^{-\beta (E_1 - E_0)} = \frac{p_1^{eq}}{p_0^{eq}},
\end{equation}
where $p^{eq}_{\lambda(t)}$ is given by the Boltzmann distribution
\begin{equation}
    p^{eq}_i = \frac{e^{-\beta E_i(\lambda)}}{Z(\lambda)},
\end{equation}
where from here on we return $\beta = 1$.

Here we can consider a protocol approximating an erasure protocol, where rather than fixing the final probabilities, we leave the final probability free and minimize work with respect to a protocol which reaches a fixed final control parameter. We focus on minimization with respect to a fixed final control rather than a fixed probability because response based approximations focus on minimization with respect to control parameters rather than the underlying probability distribution. We assume a fixed barrier energy $\Phi$, and that the initial and final control parameters are given by $\upsilon(0) = 0$ and $\upsilon(t_f) = \upsilon_f$. This setup is analogous to incomplete bit erasure (lacking restriction on the final non-equilibrium distribution before relaxation), where the probability of the erasure being effective is given by $1-\varepsilon_\upsilon$, where the control parameter required to obtain the equilibrium distribution with this error is given by $\upsilon = \ln[(1-\varepsilon_\upsilon)/\varepsilon_\upsilon]$.

\subsection{Exact Optimal Finite-time Bit Erasure}

\label{app:exact_two_state}

We can start by finding the true optimal protocols, following the approach taken by Zulkowski et al. \cite{zulkowski2014optimal}, and adapted for our case. Since $p_1 + p_0 = 1$, we can focus purely on the dynamics of $p_1$, which following Eq.~\eqref{eq:master} and substituting $p_0 = 1-p_1$ we see
\begin{equation}
    \dot p_1 = \Gamma (1-p_1) - \Gamma e^{\upsilon} p_1,
\end{equation}
which allows us to write the energy $\upsilon$ required to obtain a certain $p_1$ and $\dot p_1$ in terms of only $p_1$ and $\dot p_1$:
\begin{equation}
    \label{eq:upsilon_wrt_t}
    \upsilon(t) = \ln \bigg[\frac{1-p_1(t) - \dot p_1(t) / \Gamma}{p_ 1(t)} \bigg].
\end{equation}
Since $E_0 = 0$, the average energy is given purely by $\langle H(t)\rangle = \upsilon(t) p_1(t)$. From this, the average work over the protocol is given by
\begin{equation}
    \langle W \rangle = \int^{t_f}_0 dt \, \dot \upsilon (t) p_1(t).
\end{equation}
Integrating by parts, we can see that the work becomes
\begin{equation}
    \langle W \rangle = \upsilon(t_f)p_1(t_f) - \upsilon(0)p_1(0) - \int^{t_f}_0 dt \, \dot p_1(t) \upsilon(t),
\end{equation}
where if we insert Eq.~\eqref{eq:upsilon_wrt_t} we obtain
\begin{equation}
    \label{eq:disc_part_work}
    \begin{aligned}
        \langle W \rangle & = \upsilon(t_f)p_1(t_f) - \upsilon(0)p_1(0) \\ & - \int^{t_f}_0 dt \, \dot p_1(t) \ln \bigg[ \frac{1-p_1(t) - \dot p_1(t) / \Gamma}{p_1(t)} \bigg].
    \end{aligned}
\end{equation}
Solving the Euler-Lagrange equation to minimize the final term in the work, we find the conserved quantity
\begin{equation}
    C = \frac{\dot p_1^2}{\Gamma ( 1-p_1) - \dot p_1},
\end{equation}
where when for the branch with decreasing $p_1$, the optimal probability velocity is
\begin{equation}
    \dot p_1^\star = \frac{-C-\sqrt{C^2 + 4\Gamma C(1-p_1)}}{2}.
\end{equation}

The corresponding optimal parameter protocol during the interior of the protocol is given by plugging this into Eq.~\ref{eq:upsilon_wrt_t}. $C$ is fixed by the protocol duration and condition on the free final probability. To write this clearly we can define 
\begin{equation}
    y = \sqrt{1 + \frac{4\Gamma(1-p_1)}{C}}
\end{equation}
such that
\begin{align}
    \dot p_1^\star = -\frac{C}{2}(1+y), \ p^\star_1(y,C) = 1 - \frac{C}{4\Gamma}(y^2 - 1).
\end{align}

For a given $C$, the initial $y_i$ is fixed by the initial probability,
\begin{equation}
    y_i(C) = \sqrt{1 + \frac{4\Gamma(1- p_1(0))}{C}}.
\end{equation}
With the final probability being free, $p^\star_f = p_1(t_f)$, or equivalently $y_f$ is fixed by the free-endpoint condition given by varying the work with respect to $p_f$. From Eq.~\eqref{eq:disc_part_work} we have
\begin{equation}
    v_f - v^\star(p^\star_f, C) + \frac{\dot p^\star_f}{[\Gamma (1 - p_f^\star) - \dot p_f^\star]} = 0,
\end{equation}
where $\dot p_f^\star = \dot p_1(p^\star_f,C)$ is the velocity of the probability just before the jump at the end of the protocol. In terms of $y$, we get the control parameter before the final jump as 
\begin{equation}
    v^\star(y_f,C) = v_f - \frac{2}{(1+y_f)}.
\end{equation}
The last constraint is given by the protocol time $t_f$.
\begin{equation}
    t_f = \int_0^{t_f} dt = \int^{p^\star_f}_{1/2} \frac{1}{\dot p_1^\star} dp_1 = \int_{y_i}^{y_f} \frac{1}{\Gamma} \frac{y}{1+y} dy.
\end{equation}
From these, the exact optimal protocol is given by the constraints
\begin{align}
    v^\star(y_f,C) & = v_f - \frac{2}{(1+y_f)}, \\ t_f = & \frac{1}{\Gamma}\bigg [ y_f - y_i - \ln \bigg ( \frac{1+ y_f}{1 + y_i} \bigg ) \bigg],
\end{align}
which determine $y_f$ and $C$. Once these are known, $p^\star_1$ and $\dot p^\star_1$ are known, which fixes $\upsilon^\star(t)$ from Eq.~\eqref{eq:upsilon_wrt_t}.

\subsection{Response Based Finite-time Bit Erasure}

We can compare the exact solution derived above with the response based finite-time approximations we discussed. First, we can evaluate the linear relaxation function which depends on the equilibrium distribution and conjugate force corresponding to the control parameter $\upsilon(t)$
\begin{align}
    p^{eq}_{\upsilon(t)} = \begin{pmatrix}
        \frac{e^{\upsilon(t)}}{1+e^{\upsilon(t)}}\\
        \frac{1}{1+e^{\upsilon(t)}}
    \end{pmatrix}, \ \delta f_{\upsilon(t)} = \begin{pmatrix}
        \frac{1}{1+e^{\upsilon(t)}} \\
        \frac{-e^{\upsilon(t)}}{1+e^{\upsilon(t)}}
    \end{pmatrix}.
\end{align}
To evaluate the response functions, it will be useful to notice that the eigenvectors and eigenvalues of $K_{\lambda(t)}$ are given by
\begin{equation}
    p_{r, 0} = p^{eq}_{\upsilon(t)}, \ \ \ K_{\lambda(t)} p_{r,0} = 0
\end{equation}

\begin{equation}
    p_{r,1} = \begin{pmatrix}
        1\\
        -1
    \end{pmatrix}, \ \ \ K_{\lambda(t)} p_{r,1} = -\Gamma[1+e^{\upsilon(t)}] p_{r,1}
\end{equation}

\begin{figure*}[t]
    \centering
    \includegraphics[width=0.87\linewidth]{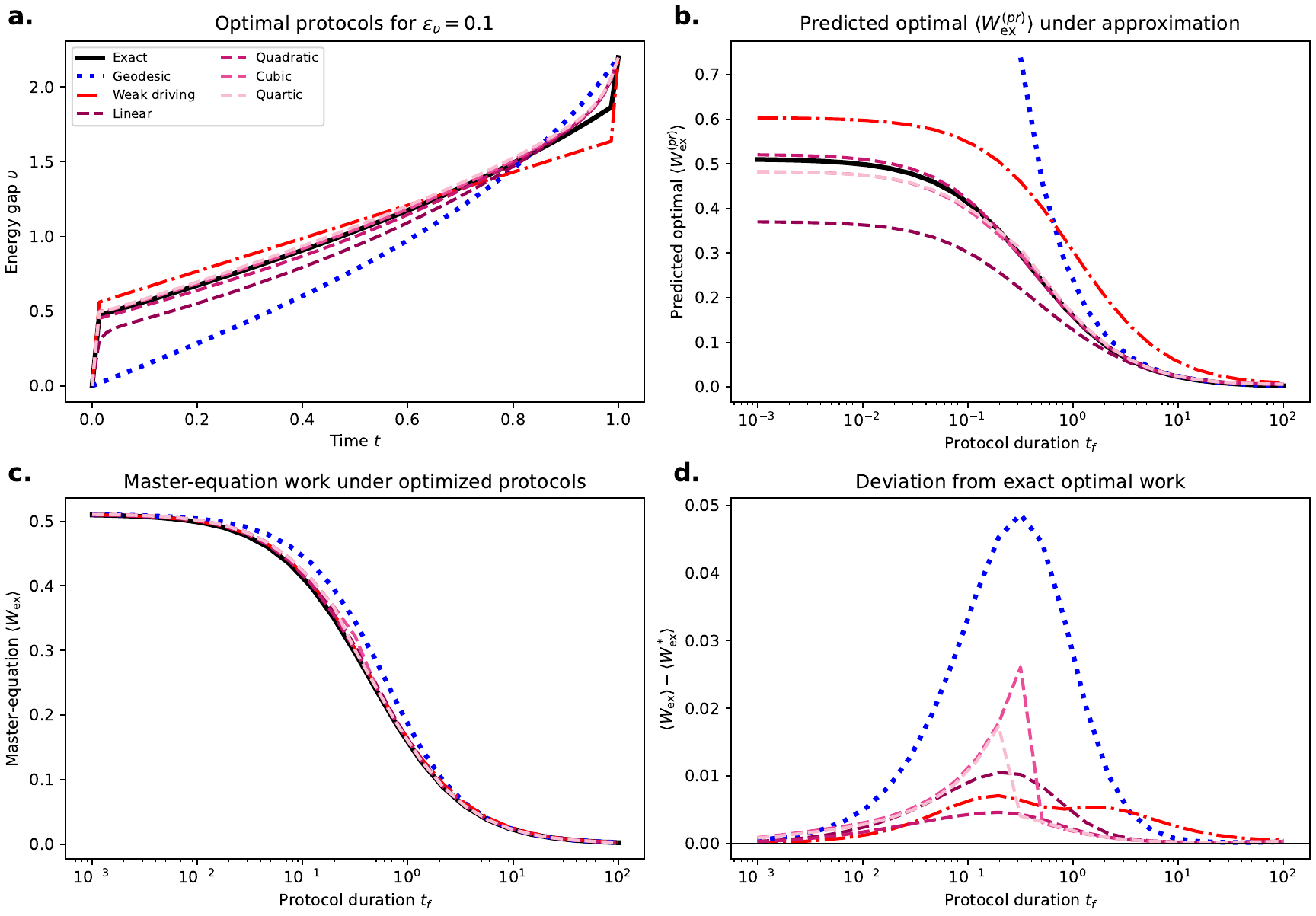}
    \caption{\textit{Comparison of different approximations of two-state optimal protocols.} Solid black = true optimal protocol from Sec.~\ref{app:exact_two_state}, blue dotted = Sivak-Crooks geodesic optimal protocol  [Eq.~\eqref{eq:two_state_geo}], red dot-dash = Bonança-Deffner weak-driving optimal protocol [Eq.~\eqref{eq:weak_lin_disc_work}], purple dashed = unconstrained linear response optimal protocol [Eq.~\eqref{eq:linear_disc_work}], lighter purple / pink dashed = unconstrained higher order response optimal protocols [Eq.~\eqref{eq:nth_stiff_work}]. All protocols are computed for a time discretization of 75 steps changing the energy level of one state from $\upsilon = 0$ to $\upsilon = 2.20$ ($\varepsilon_\upsilon = 0.1$, where $\upsilon = \ln[(1-\varepsilon_\upsilon)/\varepsilon_\upsilon]$). a. Optimal protocols for $t_f=1$ according to the different approximations. b. For different protocol durations, the predicted optimal excess work. Protocols are minimized according to the approximations from Eqs.~\eqref{eq:linear_disc_work},~\eqref{eq:weak_lin_disc_work},~\eqref{eq:two_state_geo}, and 
   ~\eqref{eq:nth_stiff_work}, and predicted work $\langle W_\text{ex}^{(pr)}\rangle$ is evaluated according to those same approximations. c. Actual excess work $\langle W_\text{ex}\rangle$ for optimal protocols suggested by each approximation at a given protocol durations, evaluated by numerically solving the master equation. d. Deviation of the true excess work for approximated optimal protocols from numerically evaluated true excess work from the exact optimal protocol $\langle W_\text{ex}^*\rangle$, as evaluated by numerically solving the master equation. The breakdown in cubic and higher order approximations (around $t_f \approx 0.4$) can be understood as due to the negativity of the cubic relaxation function at the start of the protocol [Eq.~\eqref{eq:negative-Psi}].}
    \label{fig:two-state-performance}
\end{figure*} 

\subsubsection{Linear response approximations}
The linear relaxation function is given by Eq.~\eqref{eq:linear-relaxation}. In discrete form it is
\begin{equation} 
    \begin{aligned}
        \Psi_{a b}^{(1)}(\lambda(t); \tau) & = \langle \delta f_a(0) \delta f_b(\tau) \rangle^\eq_{\lambda(t)} \\
        & = \sum_{x\in\{0,1\}} \delta f_a(x) e^{K_{\lambda(t)} \tau} [\delta f_b(x) \peq(x)].
    \end{aligned}
\end{equation}
To compute this, we find
\begin{equation}
    \delta f_{\upsilon} \odot \rho_{\upsilon(t)}^{eq} = \frac{e^{\upsilon(t)}}{(1+e^{\upsilon(t)})^2}\begin{pmatrix}
        1\\-1
    \end{pmatrix},
\end{equation}
which is an eigenvector of $K_{\upsilon(t)}$ with eigenvalue $-\Gamma(1+e^{\upsilon(t)})$. This implies that 
\begin{equation}
    e^{K_{\lambda(t)}\tau} \Big ( \delta f_b \odot \peq \Big ) = \frac{e^{\upsilon(t)}}{(1+e^{\upsilon(t)})^2}\begin{pmatrix}
        1\\-1
    \end{pmatrix} \cdot e^{-\Gamma(1+e^{\upsilon(t)}) \tau}.
\end{equation}
Finally, computing the dot product we find 
\begin{equation}
    \label{eq:linear_disc_relaxation}
    \Psi^{(1)}_{\upsilon \upsilon}(\upsilon(t); \tau) = \frac{e^{\upsilon(t)}}{(1+e^{\upsilon(t)})^2} \cdot e^{-\Gamma(1+e^{\upsilon(t)})\tau}.
\end{equation}
This implies that for unrestricted linear response, we have
\begin{equation}
    \label{eq:linear_disc_work}
    \langle W^{(1)}_\mathrm{ex} \rangle = \int^{t_f}_{0} \dt \, \dot{\upsilon}(t)\int^{t}_{0} \dt_1 \dot{\upsilon}(t_1) \Bigg[ \frac{e^{\upsilon(t)-\Gamma(1+e^{\upsilon(t)}) \cdot (t-t_1)} }{(1+e^{\upsilon(t)})^2} \Bigg].
\end{equation}
In the weak driving limit, we assume $\upsilon(t) \approx 0$, meaning
\begin{equation}
    \label{eq:weak_lin_disc_work}
    \langle W^{(1)}_\mathrm{ex, \, weak} \rangle = \int^{t_f}_{0} \dt \, \dot{\upsilon}(t)\int^{t}_{0} \dt_1 \dot{\upsilon}(t_1) \Big[ \frac{e^{-2 \Gamma \cdot (t-t_1)}}{4} \Big].
\end{equation}
Both Eq.~\ref{eq:linear_disc_work} and Eq.~\ref{eq:weak_lin_disc_work} are optimized according to the procedure described in App.~\ref{app:numerical}.

Integrating over Eq.~\eqref{eq:linear_disc_relaxation}, we see the slow driving linear response friction coefficient is given by
\begin{equation}
    \zeta_{\upsilon\upsilon} = \frac{e^{\upsilon(t)}}{\Gamma(1+e^{\upsilon(t)})^3}.
\end{equation}
Taylor expanding the velocity following \cite{sivak2012} we have
\begin{equation}
    \label{eq:geo_disc_work}
    \langle W^{(1),\text{tl}}_\mathrm{ex} \rangle = \int^{t_f}_{0} \dt \, \dot{\upsilon}^2(t) \Big[ \frac{e^{\upsilon(t)}}{\Gamma(1+e^{\upsilon(t)})^3} \Big],
\end{equation}
which can be minimized to find, for $\upsilon(0) = 0$ and $\upsilon(t_f) = \upsilon_f$,
\begin{equation}
    \label{eq:two_state_geo}
    \upsilon_{geo}(t) = \ln \bigg [ \frac{z(t)^2}{1-z(t)^2} \bigg],
\end{equation}
where
\begin{equation}
    z(t)
    =
    \frac{1}{\sqrt{2}}
    +
    \frac{t}{t_f}
    \left[
    \frac{e^{v_f/2}}{\sqrt{1+e^{v_f}}}
    -
    \frac{1}{\sqrt{2}}
    \right].
\end{equation}
This corresponds to the minimum work expression
\begin{equation}
    \langle W^{(1),\text{tl}}_\mathrm{ex,geo} \rangle =
    \frac{4}{\Gamma t_f}
    \left[
    \frac{e^{v_f/2}-1}{\sqrt{1+e^{v_f}}} - \frac{1}{\sqrt{2}}
    \right]^2 .
\end{equation}

\subsubsection{Higher-order response}

For $n^{th}$ order response, following Eq.~\ref{eq:nth-relaxation-expanded} with $0<\tau_1 < ... <  \tau_n$, we have
\begin{widetext}    
\begin{equation}
    \label{eq:nth-relaxation-discrete}
    \begin{aligned}
        \Psi_{i_0i_1\cdots i_n}^{(n)}
\bigl(\lambda(t);\tau_1,\ldots,\tau_n\bigr)
\nonumber & =
\frac{(-1)^{n+1}}{n!}
\sum_{x\in\{0,1\}}
\delta f_{i_0}(x)\\ & \cdot
\Bigg[
e^{K_{\lambda(t)}\tau_1}
\frac{\partial}{\partial\lambda^{i_1}}
\Bigg[
e^{K_{\lambda(t)}(\tau_2-\tau_1)}
\frac{\partial}{\partial\lambda^{i_2}}
\Bigg[
\cdots
\frac{\partial}{\partial\lambda^{i_{n-1}}}
\Bigg[e^{K_{\lambda(t)}(\tau_n-\tau_{n-1})}\left(\delta f_{i_n}(x) p_{\lambda(t)}^{\mathrm{eq}}(x) \right)\bigg]...\bigg]\bigg]\bigg]
    \end{aligned}
\end{equation}
\end{widetext}
Similarly to the harmonic case, this expression can be evaluated by expanding $\delta f_{i_n}(x) p^{eq}_{\lambda(t)}(x)$ in terms of the eigenvectors of $K_{\lambda(t)}$, applying the time evolution operator to the resulting expression, then taking its derivative with respect to $\lambda^{i_{n-1}}$. The next term can be evaluated by again expanding in terms of eigenvectors, applying the time evolution operator, then taking the derivative with respect to the next control parameter.

\subsubsection{Numerical Results}

While the two-state case again shows relative improvement from the inclusion of higher order terms in the slow-driving regime, it also is a counterpoint to the harmonic case which is relatively well behaved even outside of the slow driving regime. The results from this are shown in Fig.~\ref{fig:two-state-performance}.
\begin{figure}
    \centering
    \includegraphics[width=0.93\linewidth]{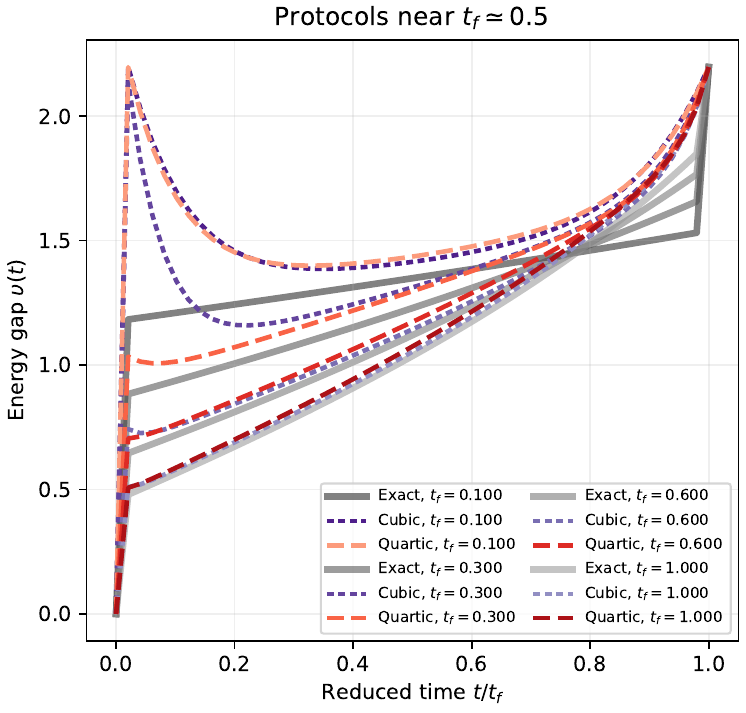}
    \caption{\textit{Cubic and quartic protocols near breakdown around $t_f \approx 0.5$.} Here we see that both the cubic and quartic approximations break down near ~$t_f \approx 0.5$, as the solution has a large jump at the start of the protocol, deviating from the exact solution (solid gray)---this is due to the negative values of the higher-order relaxation functions [e.g., Eq.~\eqref{eq:negative-Psi}]. Exact optimal protocols are shown in solid gray lines, while cubic are in dotted lines in shades of purple and quartic are dashed lines in red. 
    }
    \label{fig:cubic_quartic_spike}
\end{figure}
The instability at shorter times (around $t_f \approx 0.5$ in Fig.~\ref{fig:two-state-performance}.d) is due to the negativity of the cubic response function at the start of the protocol, which introduces an artificial method of work extraction which has an increasingly large effect at short-time protocols. By contrast, all relaxation functions are positive in the harmonic case as $\alpha \geq 0$, and $(\tau_{j+1} - \tau_j) \geq 0$, which implies overall positivity of the cost function for a monotonic protocol. The negativity of ordered cubic relaxation function can be observed by noticing that for $\upsilon = 0$ $(x = e^\upsilon = 1)$ and $0 \approx \tau_1 \approx \tau_2 \approx \tau_3$,
\begin{equation}
    \Psi^{(3), \text{ord}}_{\upsilon\upsilon\upsilon} = -\frac{1}{8}. \label{eq:negative-Psi}
\end{equation}
Because of this, once the cubic term starts to dominate at short times this term will play a larger role in the cost. The negativity of this term then allows for the optimization of unphysical negative excess work, which drives the protocol away from the true optimum as shown by Fig.~\ref{fig:two-state-performance}.d. Inclusion of the quartic term corrects this effect to an extent, however, its inclusion bears a large computational expense. Protocols near the breakdown are shown in Fig.~\ref{fig:cubic_quartic_spike}. The computation times for the two-state discrete system are shown in Tab.~\ref{tab:comp_times_disc}, showing a similar relationship to the harmonic case.

\begin{table}[htbp]
    \centering 
    \begin{tabular}{|c|c|c|c|c|} 
        \hline
        \multicolumn{5}{|c|}{\textbf{Computation times (seconds) - Two-state}} \\
        \multicolumn{5}{|c|}{single work evaluation, total optimization time} \\
        \hline
        \toprule Steps ($J$) & Linear & Quadratic & Cubic & Quartic \\ 
        % \hline
        30 & $10^{-6}$, $10^{-2}$ & $10^{-5}$, $10^{-2}$ & $10^{-4}$, $10^{-1}$ & $10^{-3}$, $10^{0}$ \\ 
        50 & $10^{-5}$, $10^{-2}$ & $10^{-4}$, $10^{-1}$ & $10^{-3}$, $10^{1}$ & $10^{-1}$, $10^{2}$ \\ 
        75 & $10^{-5}$, $10^{-1}$ & $10^{-3}$, $10^{0}$ & $10^{-2}$, $10^{1}$ & $10^{0}$, $10^{3}$ \\
        100& $10^{-4}$, $10^{-1}$ & $10^{-3}$, $10^{0}$ & $10^{-2}$, $10^{2}$ & $10^{0}$, - \\
        150& $10^{-4}$, $10^{0}$ & $10^{-3}$, $10^{1}$ & $10^{-1}$, $10^{3}$ & $10^{1}$, - \\
        \hline
    \end{tabular}
    \caption{\textit{Order-of-magnitude computation times for work evaluations and optimization for a two-state system.} Computation times shown for discretized protocols of length 30-150, with entries marked ``-" being prohibitively 
    expensive. Optimization time is computed by initializing the control protocol as linear between initial and final control parameters, and minimizing the expression given in App.~\ref{app:numerical}. Times are from evaluation on a 2025 MacBook Air laptop with an Apple M4 processor and 24 GB of RAM.}
    \label{tab:comp_times_disc} 
\end{table}

\section{Automatic Computation of Relaxation Functions}
\label{app:relaxation_functions}

As the response order increases, relaxation functions become increasingly complex. Because of this, automatic computation is helpful. Here, we discuss how automatic computation of the relaxation functions was performed for the stiffening-harmonic and two-state cases, and review the form of the larger relaxation functions. The superscript ``ord" is intended to denote that these relaxation functions are assumed to be with respect to the ordered region $0 < \tau_1 < ... < \tau_n$, meaning the symmetry factor $n!$ is already included. Recursive algebraic manipulations were performed by the \texttt{SymPy} package \cite{sympy}. Code found at: 
\url{https://github.com/SamuelDAmbrosia/Higher-order-stochastic-thermo}. 

\subsection{Stiffening Harmonic}

For the stiffening harmonic case, the identities used to derive the higher order relaxation functions are the action of the time-evolution operator on the right eigenfunction,
\begin{equation}
    e^{\mathcal{L}_{\alpha} \tau} \rho_{r,m} = e^{-m \alpha \tau} \rho_{r,m},
\end{equation}
and the identity
\begin{equation}
    \partial_\alpha \rho_{r,m} = -\frac{m}{2\alpha} \rho_{r,m} - \frac{\sqrt{(m+1)(m+2)}}{2\alpha} \rho_{r,m+2}.
\end{equation}
Since differentiation can only raise the Hermite order by 2 or leave it the same, and the excess conjugate force is given by
\begin{equation}
    \delta f_\alpha = - \frac{1}{\alpha\sqrt{2}} \rho_{l,2},
\end{equation}
the identity applied after expanding the recursive series,
\begin{equation}
    \langle \delta f_\alpha, \rho_{r,m} \rangle = -\frac{\delta_{2,m}}{\alpha \sqrt{2}},
\end{equation}
eliminates many terms (where $\delta_{2,m}$ is the Kronecker delta). This results in the following expressions, with $d_j = (\tau_{j+1} - \tau_j)$ for compactness:

\begin{align}
\Psi_{\alpha}^{(1), \text{ord}} &= \frac{e^{- 2 \alpha \tau_{1}}}{\alpha^{2}}\left[\frac{1}{2}\right] \\
\Psi_{\alpha \alpha}^{(2), \text{ord}} &= \frac{e^{- 2 \alpha \tau_{2}}}{\alpha^{3}}\left[\alpha d_{1} + 1\right] \\
\Psi_{\alpha \alpha \alpha}^{(3), \text{ord}} &= \frac{e^{- 2 \alpha \tau_{3}}}{\alpha^{4}}\left[2 \alpha^{2} d_{1} d_{2} + 2 \alpha^{2} d_{2}^{2} + 2 \alpha d_{1} + 4 \alpha d_{2} + 3\right]
\end{align}
\begin{equation}
    \begin{aligned}
        \Psi_{\alpha \alpha \alpha \alpha}^{(4), \text{ord}} & = \frac{e^{- 2 \alpha \tau_{4}}}{\alpha^{5}} [2 (2 \alpha^{3} d_{1} d_{2} d_{3} + 2 \alpha^{3} d_{1} d_{3}^{2} \\ & + 2 \alpha^{3} d_{2}^{2} d_{3} + 4 \alpha^{3} d_{2} d_{3}^{2} + 2 \alpha^{3} d_{3}^{3} + 2 \alpha^{2} d_{1} d_{2} \\ & + 4 \alpha^{2} d_{1} d_{3} + 2 \alpha^{2} d_{2}^{2} + 8 \alpha^{2} d_{2} d_{3} \\ & + 6 \alpha^{2} d_{3}^{2} + 3 \alpha d_{1} + 6 \alpha d_{2} + 9 \alpha d_{3} + 6)]
    \end{aligned}
\end{equation}

\subsection{Two-state}

For the two-state case, a similar set of steps is used to derive the higher order relaxation functions based on recursively applying the time evolution operator then taking a derivative. A helpful simplification is using $x = e^{\upsilon}$, which avoids cumbersome expressions by providing
\begin{equation}
    \partial_\upsilon = x \partial_x.
\end{equation}
While the Hermite identities or orthogonality relations are not applicable here, the size of the state space keeps the expressions from getting too big. The recursion results in the following expressions, with $x = e^{\upsilon}$ and again $d_j = (\tau_{j+1} - \tau_j)$ for compactness:

\begin{align}
\Psi_{\upsilon}^{(1), \text{ord}} &= \frac{x e^{- \Gamma \tau_{1} \left(x + 1\right)}}{\left(x + 1\right)^{2}}\left[1\right] \\
\Psi_{\upsilon\upsilon}^{(2), \text{ord}} &= \frac{x e^{- \Gamma \tau_{2} \left(x + 1\right)}}{\left(x + 1\right)^{3}}\left[\Gamma d_{1} x \left(x + 1\right) + \left(x - 1\right)\right]
\end{align}
\begin{equation}
    \begin{aligned}
        \Psi_{\upsilon\upsilon\upsilon}^{(3), \text{ord}} &= \frac{x e^{- \Gamma \tau_{3} (x + 1)}}{(x + 1)^{4}}[\Gamma^{2} d_{2} x^{2} (d_{1} + d_{2}) (x + 1)^{2} \\ & + \Gamma x (x + 1) (d_{1} x - d_{1} + d_{2} x - 3 d_{2}) + (x^{2} - 4 x + 1)]
    \end{aligned}
\end{equation}
\begin{equation}
    \begin{aligned}
        \Psi_{\upsilon\upsilon\upsilon\upsilon}^{(4), \text{ord}} &= \frac{x e^{- \Gamma \tau_{4} (x + 1)}}{(x + 1)^{5}}[\Gamma^{3} d_{3} x^{3} (d_{2} + d_{3}) (x + 1)^{3} (d_{1} + d_{2} + d_{3})\\ & + \Gamma^{2} x^{2} (x + 1)^{2} (d_{1} d_{2} x - d_{1} d_{2} + d_{1} d_{3} x - 3 d_{1} d_{3} + d_{2}^{2} x \\ &- d_{2}^{2} + d_{2} d_{3} x - 7 d_{2} d_{3} - 6 d_{3}^{2}) + \Gamma x (x + 1) (d_{1} x^{2} \\ &- 4 d_{1} x + d_{1} + d_{2} x^{2} - 8 d_{2} x + 3 d_{2} + d_{3} x^{2} \\ & - 10 d_{3} x + 7 d_{3}) + (x - 1) (x^{2} - 10 x + 1)]
    \end{aligned}
\end{equation}

\appendix

\bibliography{apssamp}

\end{document}